\documentclass[american,english]{article}
\usepackage[T1]{fontenc}
\usepackage[latin9]{inputenc}
\usepackage{mathrsfs}
\usepackage{amsbsy}
\usepackage{amssymb}
\usepackage{cancel}
\usepackage{esint}

\makeatletter
\newcommand{\lyxaddress}[1]{
	\par {\raggedright #1
	\vspace{1.4em}
	\noindent\par}
}

\usepackage{fullpage}

\makeatother

\usepackage{babel}
\begin{document}
\title{Dirac Sources for Nonmetricity and Torsion in Metric-affine Gravity}
\author{James T. Wheeler$^{\dagger}$}
\maketitle
\begin{abstract}
Metric-affine gravity (GL(4) gauge theory) in 4-dimensions is coupled
to a spacetime Dirac source field using the isomorphisms of the Lie
algebra gl(4) to the Clifford algebras Cl(3,1) and Cl(2,2). A simple
transformation relates the generators of Cl(3,1) to a real representation
of Cl(2,2), while the real representation of Cl(2,2) serves directly
as a basis for the Lie algebra gl(4). Therefore, although GL(4) does
not contain a spinor representation of the Lorentz group, expanding
its Lie algebra in the Cl(2,2) basis gives a Clifford valued connection
with well-defined coupling to Dirac spinors. Variation of the expansion
coefficients gives new Dirac sources for both torsion and nonmetricity,
separated by identifying the so$\left(3,1\right)$ basis within the
gl(4) basis.
\end{abstract}

\lyxaddress{Keywords: General linear gauge theory, GL(4) gravity, Poincaré gauge
theory, torsion, nonmetricity, non-metricity, nonmetric gravity, non-metric
gravity, general relativity, metric-affine gravity, hypermomentum,
particle-antiparticle asymmetry}

$^{\dagger}$James T Wheeler, Utah State University Department of
Physics, 4415 Old Main Hill, Logan, UT 84322-4415, jim.wheeler@usu.edu

\newpage{}

\section{Introduction}

When the metric and connection on a 4-dimensional manifold $\left(\mathcal{M},g,\Sigma\right)$
are treated independently, the connection may be divided by symmetry
into nonmetricity and torsion, with the symmetric nonmetric part first
explored by H. Weyl \cite{Weyl 1918,Weyl 1918a,Weyl 1918b,Weyl 1919}
and the antisymmetric torsion part introduced by É. Cartan \cite{Cartan 1922 intro of torsion,Cartan,Cartan 1922,Cartan 1923,Cartan 1924}.
These tensors may both be defined as covariant exterior derivatives
\begin{eqnarray*}
\mathbf{Q}_{ab} & = & \boldsymbol{\mathfrak{D}}g_{ab}\:=\:\mathbf{d}g_{ab}-g_{cb}\boldsymbol{\Sigma}_{\;\;\;a}^{c}-g_{ac}\boldsymbol{\Sigma}_{\;\;\;b}^{c}\\
\boldsymbol{\mathfrak{T}}^{a} & = & \boldsymbol{\mathfrak{D}}\mathbf{e}^{a}\:=\:\mathbf{d}\mathbf{e}^{a}-\mathbf{e}^{b}\land\boldsymbol{\Sigma}_{\;\;\;b}^{a}
\end{eqnarray*}
where the basis 1-form $\mathbf{e}^{a}$ and the metric $g_{ab}$
are related by the inner product $\left\langle \mathbf{e}^{a},\mathbf{e}^{b}\right\rangle =g^{ab}$
with $g^{ab}g_{bc}=\delta_{c}^{a}$. Here $\boldsymbol{\mathfrak{D}}$
is the covariant exterior derivative with connection 1-form $\boldsymbol{\Sigma}_{\;\;\;b}^{a}$,
the 2-form $\boldsymbol{\mathfrak{T}}^{a}$ is the torsion, and the
1-form $\mathbf{Q}_{ab}$ is the nonmetricity. If the basis $\mathbf{e}^{a}$
is orthonormal $\left\langle \mathbf{e}^{a},\mathbf{e}^{b}\right\rangle =\eta^{ab}$
the nonmetricity becomes algebraic
\begin{eqnarray*}
\mathbf{Q}_{ab} & = & \cancel{\mathbf{d}\eta_{ab}}-\eta_{cb}\boldsymbol{\Sigma}_{\;\;\;a}^{c}-\eta_{ac}\boldsymbol{\Sigma}_{\;\;\;b}^{c}=\:-\left(\boldsymbol{\Sigma}_{ba}+\boldsymbol{\Sigma}_{ab}\right)
\end{eqnarray*}
Alternatively, if we use a coordinate basis $\mathbf{e}^{a}=\delta_{\mu}^{\;\;\;a}\mathbf{d}x^{\mu}$
then the torsion becomes algebraic
\begin{eqnarray*}
\boldsymbol{\mathfrak{T}}^{a} & = & \cancel{\mathbf{d}^{2}x^{a}}-\mathbf{e}^{b}\land\boldsymbol{\Sigma}_{\;\;\;b}^{a}\:=-\Sigma_{\;\;\;\mu\nu}^{a}\mathbf{d}x^{\mu}\land\mathbf{d}x^{\nu}
\end{eqnarray*}
so that $\mathfrak{T}_{\;\;\;\mu\nu}^{a}=\Sigma_{\;\;\;\nu\mu}^{a}-\Sigma_{\;\;\;\mu\nu}^{a}$.
Here we use Latin indices $\left(a,b\ldots=0,1,2,3\right)$ for orthonormal
frames and Greek indices $\left(\mu,\nu\ldots=0,1,2,3\right)$ for
coordinate frames.

Torsion arises naturally within ECSK gravity, including Cartan's introductory
work \cite{Cartan 1922 intro of torsion,Cartan,Cartan 1922,Cartan 1923,Cartan 1924}
together with \cite{Einstein 1928,Kibble 1961,Sciama1962,Sciama,HehlDatta,Datta,Hehl,HehlvonderHeydeKerlick,HehlNester}.
This may be understood as the gauge theory of the Poincaré group using
the Einstein-Hilbert form of the action but with an asymmetric connection.
In ECSK gravity the curvature is the field strength of the Lorentz
connection and the torsion is the field strength of the gauge field
of translations.

Nonmetric geometries include Weyl geometry (see review in \cite{Wheeler2018}),
which depends on a single vector given by the trace of the nonmetricity.
Geometries with general nonmetricity have been largely confined to
metric-affine gravity \cite{HehlKerlickVonDerHeyde,Hehl et al III,HehlLordNe'eman,HehlNitschvonderHeyde,HehlMetricAffine,RigouzzoZell},
the gauge theory of the general linear group $GL\left(4,\mathbb{R}\right)$
\footnote{But see \cite{Wheeler2025a}, showing a relationship between nonmetricity
and the conformal curvatures.}. The metric and connection are necessarily treated independently
since the general linear group does not single out a metric.

Although neither torsion nor nonmetricity has any direct experimental
support, the study of these more general geometries gives a broader
arena for tests of general relativity, and may lead to their ultimate
measurement or explain why we do not see them. Of particular interest
in this regard are possible Standard Model sources. As we show in
detail in the next Section, neither torsion nor nonmetricity is driven
by Yang-Mills gauge theories or scalars, because the actions of those
fields do not depend on the spacetime connection. Dirac and Rarita-Schwinger
spinor fields, however, do provide sources for torsion in ECSK gravity
\cite{Wheeler2023a}.

The goal for our current work is to identify Dirac sources for both
torsion and nonmetricity in $GL\left(4\right)$ gravity. It is important
to observe that since the sources for torsion and nonmetricity both
depend on the overall symmetry of the connection, the source for torsion
differs between ECSK and $GL\left(4\right)$ (metric-affine) gravity.
Concretely, Poincaré gauge theory couples only the totally antisymmetric
part of the torsion to a single Dirac pseudo-current, but as we show
here, all 16 Dirac currents play a role in metric-affine gravity.
The $GL\left(4\right)$ sources for torsion differ strongly from the
predictions of ECSK theory, while the full range of Dirac sources
for generic nonmetricity have not previously been identified.

The principal issue encountered with Dirac sources in metric-affine
gravity is that the general linear group $GL\left(4,\mathbb{R}\right)$
does not have finite dimensional spinor representations. Early studies
of metric-affine gravity \cite{HehlKerlickVonDerHeyde} noted that
the source for the metric-affine connection, called \emph{hypermomentum},
does not easily include spinors:
\begin{quotation}
``This last remark {[}that the traceless nonmetricity depends on
generators of general linear transformations{]} seems to preclude
the definition of a canonical hypermomentum tensor for spinor fields,
since $GL\left(4,\mathbb{R}\right)$ has no spinor representations.
For the time being, at least, we exclude spinor fields from our consideration
. . ..''
\end{quotation}
Since the Yang-Mills and scalar fields of the standard model do not
provide sources for nonmetricity, excluding spinor sources ignores
direct coupling of nonmetricity to the Standard Model. Nonetheless,
coupling is possible because the connection depends only on the \emph{Lie
algebra} $\mathfrak{gl}\left(4\right)$, which \emph{does} have spinor
representations. An interesting recent study \cite{RigouzzoZell}
includes Dirac sources, but unlike our presentation \cite{RigouzzoZell}
restricts the fermion coupling to the antisymmetric part of the connection.
This restriction means that \cite{RigouzzoZell} only couples spinors
to traces of the torsion and nonmetricity. Our current work couples
the full connection to spinors.

\emph{Our central result is to find sources for nonmetricity and torsion
in metric-affine gravity. This first requires showing how to couple
spinors to the full general linear connection, despite the lack of
spinor representations of $GL\left(4\right)$}. The key to accomplishing
this to use the isomorphism between the \emph{Lie algebras}--$\mathfrak{gl}$$\left(4,\mathbb{R}\right)$
for the general linear group and the Clifford algebras $\mathfrak{Cl}\left(3,1\right)$
and $\mathfrak{Cl}\left(2,2\right)$ for spinors. Since each of these
algebras span all $4\times4$ matrices the infinitesimal action of
the general linear group on spinors is well-defined. Writing the elements
of $\mathfrak{gl}$$\left(4,\mathbb{R}\right)$ in the Clifford basis,
the sources for nonmetricity are straightforward (but lengthy) to
compute.

A second seeming obstacle to finding spinor sources in $GL\left(4\right)$
gravity is the definition of the spinors themselves, because spinors
are defined as representations of a Clifford algebra. The Clifford
algebra, in turn, depends on the Lorentz metric $\eta_{ab}$ through
the definition of the gamma matrices, $\left\{ \gamma^{a},\gamma^{b}\right\} =-2\eta^{ab}1$.
The resolution lies in the independence of metric and connection in
metric-affine gravity, which allows us to \emph{choos}e the Minkowski
metric. Since any metric we choose will be \emph{not} be compatible
with the connection, we have this freedom.

\medskip{}

A class of sources for the nonmetricity in metric-affine gravity (the
gauge theory of the general linear group), is defined by the variation
of the action with respect to the connection \cite{HehlNester} 
\[
e\Delta_{a}^{\;\;\;bc}\equiv-\frac{\delta\mathcal{L}_{source}}{\delta\Sigma_{\;\;\;bc}^{a}}
\]
where the metric and connection are independent. The source tensor
$\Delta_{a}^{\;\;\;bc}$ is called the \emph{hypermomentum}. By expanding
the $GL\left(4\right)$ connection $\Sigma_{\;\;\;bc}^{a}$ in a $\mathfrak{cl}\left(3,1\right)$
basis we can describe the hypermomentum of the spinor field. The process
is simplified by working with the Clifford algebra $\mathfrak{Cl}\left(2,2\right)$
built from a real basis of gamma matrices for $\mathfrak{spin}\left(2,2\right)$.
The resulting real basis for $\mathfrak{Cl}\left(2,2\right)$ serves
directly as a basis for $\mathfrak{gl}\left(4\right)$. At the same
time there is a simple transformation from that real basis to a basis
for $\mathfrak{Cl}\left(3,1\right)$.

\medskip{}

We use the Einstein-Hilbert form of the gravitational action, but
build the curvature from the general linear connection, leading to
both torsion and nonmetric dependence. Then, to separate the torsion
and nonmetric dependence of the action, we identify the $SO\left(3,1\right)$
subgroup of $GL\left(4\right)$. The remaining dependence is identified
with nonmetricity. For sources we consider the three types of matter
field within the Standard Model: the Higgs scalar doublet, $U\left(1\right)$
and Yang-Mills fields, and Dirac spinors. We show in the next Section
that only Dirac fields provide sources \footnote{It is easy to see that non-gauge tensor fields can provide sources.
Let $S_{T}=\int\left(D^{\alpha}V^{\mu\ldots\nu}D_{\alpha}V_{\mu\ldots\nu}+m^{2}V^{\mu\ldots\nu}V_{\mu\ldots\nu}\right)\sqrt{-g}d^{4}x$
and vary $H_{\beta\mu\nu}=-\frac{1}{2}\left(Q_{\beta\mu\nu}+Q_{\beta\nu\mu}-Q_{\mu\nu\beta}\right)$.
With $\delta_{H}\left(D_{\alpha}V_{\mu\ldots\nu}\right)=-V_{\beta\ldots\nu}\delta H_{\;\;\;\mu\alpha}^{\beta}-\cdots-V_{\mu\ldots\beta}\delta H_{\;\;\;\nu\alpha}^{\beta}$we
have a coupling $\delta_{H}S_{T}=-2\int D^{\alpha}V^{\mu\ldots\nu}\left(V_{\beta\ldots\nu}\delta_{\mu}^{\rho}+\cdots+V_{\mu\ldots\beta}\delta_{\nu}^{\rho}\right)\delta H_{\;\;\;\rho\alpha}^{\beta}\sqrt{-g}d^{4}x$
so that the nonmetricity has a source. This dependence vanishes with
antisymmetrization so that Yang-Mills fields depend only on the internal
$SU\left(N\right)$ connection.}.

\medskip{}

In Section \ref{sec:Einstein-Hilbert-action-with} we expand the metric-affine
curvature, separating the torsion and nonmetric dependence from the
usual Riemannian curvature, then vary to find the gravitational contribution
to the torsion and nonmetric field equations. We show the well-known
result that in vacuum the torsion vanishes. The nonmetric variation
shows that only the totally symmetric part of the nonmetricity vanishes.
The remaining part of the nonmetricity consists of the Weyl vector
and an antisymmetric piece which may be absorbed into the torsion
\cite{Wheeler2025a}. In Section \ref{sec:Clifford-basis-and} we
present details of the Clifford algebras $\mathfrak{Cl}\left(3,1\right)$
and $\mathfrak{Cl}\left(2,2\right)$, then write the coupled form
of the Dirac action. In Section \ref{sec:Nonmetricity-vs.-torsion:}
we separate the generators of the $SO\left(3,1\right)$ subgroup of
$GL\left(4\right)$, allowing us to identify the part of the connection
variation relating to torsion. Our main results, the Dirac sources
for torsion and nonmetricity in metric-affine gravity, are presented
in Section \ref{sec:The-field-equations}, with a summary and some
simple cases in the final Section.

For the orthonormal vector basis we use lower case Latin letters from
the beginning of the alphabet, $\eta_{ab},\mathbf{e}^{a}\left(a,b,\ldots=0,1,2,3\right)$,
with indices $i,j,\ldots=1,2,3$. For coordinates we use lower case
Greek, $g_{\alpha\beta},v^{\mu}$. For spinors we use upper case Latin
indices, e.g., $\psi^{A},\left[\gamma^{a}\right]_{\;\;\;C}^{B}$,
and for the Clifford basis we use upper case Greek, $\Gamma^{\Delta}\in\left\{ 1,\gamma^{a},\sigma^{ab},\gamma_{5}\gamma^{a},\gamma_{5}\right\} ,\left(\Delta,\Xi,\ldots=1,\ldots,16\right)$.
The Clifford basis may be divided into symmetric $\Gamma^{\Delta_{s}}$
and antisymmetric $\Gamma^{\Delta_{a}}$ parts. When needed we use
the Dirac form of the gamma matrices, given in \ref{sec:Clifford-basis-and}.

\section{Einstein-Hilbert action with torsion and nonmetric additions \label{sec:Einstein-Hilbert-action-with}}

\subsection{Decomposition of the connection and curvature}

The spacetime arena we consider is the class of geometries modeled
on the principle $GL\left(4\right)$ fiber bundle based on the quotient
of the affine group $A\left(4\right)=R^{4}\rtimes GL\left(4\right)$
by the general linear group, $\mathcal{M}_{0}^{4}=A\left(4\right)/GL\left(4\right)$.
Generalizing the homogeneous quotient manifold $\mathcal{M}_{0}^{4}\rightarrow\mathcal{M}^{4}$
and the flat Maurer-Cartan connection to the curved connection $\boldsymbol{\Sigma}_{\;\;\;b}^{a}$,
we have the Cartan structure equations:
\begin{eqnarray}
\mathbf{d}\mathbf{e}^{a} & = & \mathbf{e}^{b}\land\boldsymbol{\Sigma}_{\;\;\;b}^{a}+\boldsymbol{\mathfrak{T}}^{a}\label{Torsion}\\
\mathbf{d}\boldsymbol{\Sigma}_{\;\;\;b}^{a} & = & \boldsymbol{\Sigma}_{\;\;\;b}^{c}\land\boldsymbol{\Sigma}_{\;\;\;c}^{a}+\boldsymbol{\mathfrak{R}}_{\;\;\;b}^{a}\label{Curvature}
\end{eqnarray}
These define the torsion $\boldsymbol{\mathfrak{T}}^{a}$ and curvature
$\boldsymbol{\mathfrak{R}}_{\;\;\;b}^{a}$, while the nonmetricity
is defined by
\begin{eqnarray}
\mathbf{Q}_{ab}\:=\;\boldsymbol{\mathfrak{D}}g_{ab} & = & \mathbf{d}g_{ab}-g_{cb}\boldsymbol{\Sigma}_{\;\;\;a}^{c}-g_{ac}\boldsymbol{\Sigma}_{\;\;\;b}^{c}\label{Nonmetricity}
\end{eqnarray}
Keeping the fields $\boldsymbol{\mathfrak{T}}^{a}$ and $\boldsymbol{\mathfrak{R}}_{\;\;\;b}^{a}$
horizontal preserves the principal bundle structure with $GL\left(4\right)$
connection $\boldsymbol{\Sigma}_{\;\;\;b}^{a}$.

Field redefinition below expresses the full $GL\left(4\right)$ torsion
$\boldsymbol{\mathfrak{T}}^{a}$ (Fraktur script) in terms of redefined
torsion $\mathbf{T}^{a}\equiv\boldsymbol{\mathfrak{T}}^{a}-\mathbf{Q}^{a}$.
The $GL\left(4\right)$ curvature $\boldsymbol{\mathfrak{R}}_{\;\;\;b}^{a}$
is broken into the Einstein-Cartan curvature $\boldsymbol{\mathcal{R}}^{ab}$
(Calligraphic script) plus nonmetric terms, and further into the familiar
Riemann curvature of a symmetric, metric compatible connection $\mathbf{R}^{ab}$
(Roman script) plus contorsion terms.

To include Standard Model sources we extend this spacetime background
by including the Standard Model symmetries. The quotient $\mathcal{M}_{0}^{4}=\left[A\left(4\right)\times U\left(1\right)\times SU\left(2\right)\times SU\left(3\right)\right]/\left[GL\left(4\right)\times U\left(1\right)\times SU\left(2\right)\times SU\left(3\right)\right]$
is then a principal fiber bundle with base manifold $\mathcal{M}_{0}^{4}$
and fibers $GL\left(4\right)\times U\left(1\right)\times SU\left(2\right)\times SU\left(3\right)$.
In addition to Eqs.(\ref{Torsion}) and (\ref{Curvature}), the Cartan
equations now include
\begin{eqnarray}
\mathbf{d}\mathbf{A} & = & \mathbf{F}\nonumber \\
\mathbf{d}\mathbf{W}^{a} & = & -\frac{1}{2}c_{\;\;\;bc}^{a}\mathbf{W}^{b}\wedge\mathbf{W}^{c}+\mathbf{F}^{a}\quad\left(a,b,c=1,2,3\right)\nonumber \\
\mathbf{d}\mathbf{B}^{K} & = & -\frac{1}{2}f_{\;\;\;LM}^{K}\mathbf{B}^{L}\wedge\mathbf{B}^{M}+\mathbf{H}^{K}\quad\left(K,L,M=1,\ldots,8\right)\label{Yang-Mills fields}
\end{eqnarray}
where $\mathbf{A},\mathbf{W}^{a}$ are the $U\left(1\right)\times SU\left(2\right)$
electroweak fields, $\mathbf{B}^{K}$ are the gluon gauge fields,
and $c_{\;\;\;bc}^{a},f_{\;\;\;LM}^{K}$ are the $SU\left(2\right)$
and $SU\left(3\right)$ structure constants, respectively. Then Eqs.(\ref{Yang-Mills fields})
define the field strengths $\mathbf{F},\mathbf{F}^{a}$, and $\mathbf{G}^{K}$.
The essential feature is that the fields $\mathbf{F},\mathbf{F}^{a}$
and $\mathbf{G}^{K}$\emph{ do not depend on the G$\left(4\right)$
connection $\boldsymbol{\Sigma}_{\;\;\;b}^{a}$} , so varying $\boldsymbol{\Sigma}_{\;\;\;b}^{a}$
in the corresponding action $S_{YM}=\int a\mathbf{F}\wedge^{*}\mathbf{F}+b\delta_{ab}\mathbf{F}^{a}\wedge^{*}\mathbf{F}^{b}+c\delta_{KM}\mathbf{G}^{K}\wedge^{*}\mathbf{G}^{M}$
gives no contribution. \emph{As a result, within the Standard Model
only Dirac fields provide sources for torsion and nonmetricity.} For
the remainder of our discussion we have no further need of Eqs.(\ref{Yang-Mills fields}).

We next separate the contributions of torsion and nonmetricity to
the curvature. Choose the orthonormal basis, $g_{ab}=\eta_{ab}$,
so that $\boldsymbol{\Sigma}_{ba}+\boldsymbol{\Sigma}_{ab}=-\mathbf{Q}_{ab}$
and let $\boldsymbol{\Omega}_{ab}=\boldsymbol{\Sigma}_{\left[ab\right]}$.
Then the connection is
\begin{eqnarray}
\boldsymbol{\Sigma}_{ab} & = & -\frac{1}{2}\mathbf{Q}_{ab}+\boldsymbol{\Omega}_{ab}\label{Connection by symmetry}
\end{eqnarray}
Substituting Eq.(\ref{Connection by symmetry}) into Eq.(\ref{Torsion})
results in
\begin{eqnarray*}
\mathbf{d}\mathbf{e}^{a} & = & \mathbf{e}^{b}\land\boldsymbol{\Omega}_{\;\;\;b}^{a}+\boldsymbol{\mathfrak{T}}^{a}-\mathbf{Q}^{a}
\end{eqnarray*}
where we define the nonmetric 2-form $\mathbf{Q}^{a}\equiv\frac{1}{2}\mathbf{e}^{b}\land\mathbf{Q}_{\;\;\;b}^{a}$.
We may eliminate explicit dependence on the nonmetric 2-form $\mathbf{Q}^{a}$
by the field redefinition, ,
\begin{eqnarray}
\mathbf{T}^{a} & \equiv & \boldsymbol{\mathfrak{T}}^{a}-\mathbf{Q}^{a}\label{Torsion redefinition}
\end{eqnarray}
 leaving the nonmetricity determined by its remaining, totally symmetric
part $Q_{\left(abc\right)}$. The resulting structure equation determines
a Lorentz connection with torsion
\begin{eqnarray*}
\mathbf{d}\mathbf{e}^{a} & = & \mathbf{e}^{b}\land\boldsymbol{\Omega}_{\;\;\;b}^{a}+\mathbf{T}^{a}
\end{eqnarray*}

We solve for $\boldsymbol{\Omega}_{\;\;\;b}^{a}$ by defining the
torsion-free, metric-compatible spin connection $\boldsymbol{\omega}_{\;\;\;b}^{a}$
satisfying $\mathbf{d}\mathbf{e}^{a}=\mathbf{e}^{b}\land\boldsymbol{\omega}_{\;\;\;b}^{a}$.
Then setting $\boldsymbol{\Omega}_{\;\;\;b}^{a}=\boldsymbol{\omega}_{\;\;\;b}^{a}-\mathbf{C}_{\;\;\;b}^{a}$,
the contorsion $\mathbf{C}_{\;\;\;b}^{a}$ must satisfy $0=\mathbf{e}^{b}\land\mathbf{e}^{c}\left(-C_{\;\;\;bc}^{a}+\frac{1}{2}T_{\;\;\;bc}^{a}\right)$.
Lowering $a$ and cycling indices, we add the first two permutations
and subtract the third to find
\begin{eqnarray}
C_{abc} & = & \frac{1}{2}\left(T_{abc}+T_{bca}-T_{cab}\right)\label{Contorsion}
\end{eqnarray}
 We may recover the torsion as $\mathbf{e}^{b}\wedge\mathbf{C}_{\;\;\;b}^{a}=\mathbf{T}^{a}$.

With $\boldsymbol{\Omega}_{\;\;\;b}^{a}=\boldsymbol{\omega}_{\;\;\;b}^{a}-\mathbf{C}_{\;\;\;b}^{a}$,
Eq.(\ref{Connection by symmetry}) becomes $\boldsymbol{\Sigma}_{\;\;\;b}^{a}=\boldsymbol{\omega}_{\;\;\;b}^{a}-\mathbf{C}_{\;\;\;b}^{a}-\frac{1}{2}\mathbf{Q}_{\;\;\;b}^{a}$.
Substituting this into Eq.(\ref{Curvature}) results in 
\begin{eqnarray}
\boldsymbol{\mathfrak{R}}^{ab} & = & \boldsymbol{\mathcal{R}}^{ab}-\frac{1}{2}\mathbf{D}\mathbf{Q}^{ab}-\frac{1}{4}\mathbf{Q}^{cb}\land\mathbf{Q}_{\;\;\;c}^{a}-\mathbf{Q}^{c(a}\land\mathbf{C}_{\;\;\;c}^{b)}\label{Curvature with C and Q separated}
\end{eqnarray}
where $\mathbf{D}$ is the usual Lorentz covariant exterior derivative
and $\boldsymbol{\mathcal{R}}_{\;\;\;b}^{a}=\mathbf{R}_{\;\;\;b}^{a}-\mathbf{D}\mathbf{C}_{\;\;\;b}^{a}-\mathbf{C}_{\;\;\;b}^{c}\land\mathbf{C}_{\;\;\;c}^{a}$
is the Lorentz curvature with torsion.

Equations (\ref{Torsion redefinition}) and (\ref{Curvature with C and Q separated})
now express the original curvature and torsion in terms of $\mathbf{R}^{ab},\mathbf{Q}^{ab}$
and $\mathbf{T}^{a}$.

\subsection{The action}

The action functional is $S_{Grav}+S_{D}$ where $S_{Grav}$ given
by using Eq.\ref{Curvature with C and Q separated} in the Einstein-Hilbert
form of the action $S_{Grav}=\frac{\kappa}{2}\int\boldsymbol{\mathfrak{R}}^{ab}\land\mathbf{e}^{c}\land\mathbf{e}^{d}e_{abcd}$.
The sources for torsion and nonmetricity depend on the variation of
$S_{EH}$ with respect to $\boldsymbol{\Omega}_{\;\;\;b}^{a}$ and
$\mathbf{Q}_{\;\;\;c}^{a}$, respectively, Expanding, the symmetric
terms $-\frac{1}{2}\mathbf{D}\mathbf{Q}^{ab}-\mathbf{Q}^{c(a}\land\mathbf{C}_{\;\;\;c}^{b)}$
drop out when contracted with the Levi-Civita tensor, resulting in
full separation of $\boldsymbol{\Omega}^{ab}$ and $\boldsymbol{Q}^{ab}$
\begin{eqnarray}
S_{Grav}\left[\Omega,Q\right] & = & \frac{\kappa}{2}\int\left(\boldsymbol{\mathcal{R}}^{ab}-\frac{1}{4}\mathbf{Q}^{bc}\land\mathbf{Q}_{\;\;\;c}^{a}\right)\land\mathbf{e}^{c}\land\mathbf{e}^{d}e_{abcd}\label{Action}
\end{eqnarray}

Varying $\boldsymbol{\Omega}^{ab}$ within $\boldsymbol{\mathcal{R}}^{ab}$,
\begin{eqnarray*}
\delta_{\Omega}S_{Grav} & = & \kappa\int\delta\boldsymbol{\Omega}^{eb}\land\mathbf{C}_{\;\;\;e}^{a}\wedge\mathbf{e}^{c}\wedge\mathbf{e}^{d}e_{abcd}\\
 & = & \kappa\int\delta\Omega_{\quad f}^{eb}\left(C_{\;\;\;eb}^{f}-C_{\;\;\;be}^{f}-C_{\;\;\;ea}^{a}\delta_{b}^{f}+C_{\;\;\;ba}^{a}\delta_{e}^{f}\right)\boldsymbol{\Phi}
\end{eqnarray*}
where the volume form $\boldsymbol{\Phi}$ is given by the Hodge dual
of one, $^{*}1$. Using Eq.(\ref{Contorsion}) the integrand $\mathscr{T}_{\;\;\;eb}^{g}=C_{\;\;\;eb}^{g}-C_{\;\;\;be}^{g}-C_{\;\;\;ea}^{a}\delta_{b}^{g}+C_{\;\;\;ba}^{a}\delta_{e}^{g}$
reduces to a trace-altered form of the torsion
\begin{equation}
\mathscr{T}_{\;\;\;bc}^{a}=T_{\;\;\;bc}^{a}-\delta_{b}^{a}T_{\;\;\;ec}^{e}+\delta_{c}^{a}T_{\;\;\;eb}^{e}\label{ECSK source for torsion}
\end{equation}
Equate this to the source contribution,
\begin{eqnarray}
\kappa\mathscr{T}_{\;\;\;bc}^{a} & = & -\frac{\delta\mathcal{L}_{Source}}{\delta\Omega^{abc}}\label{Field eq for torsion}
\end{eqnarray}

The variation of nonmetricity is similar. We find
\begin{eqnarray*}
\delta_{Q}S_{Grav} & = & -\frac{\kappa}{4}\int\delta_{Q}Q_{\quad f}^{eb}\left(Q_{\;\;\;eb}^{f}+Q_{\;\;\;be}^{f}+Q_{\;\;\;ea}^{a}\delta_{b}^{f}+Q_{\;\;\;ba}^{a}\delta_{e}^{f}\right)\boldsymbol{\Phi}
\end{eqnarray*}
Defining the trace-altered nonmetricity,
\begin{eqnarray}
\mathcal{\mathscr{Q}}_{acb} & \equiv & -\frac{1}{4}\left(Q_{acb}+Q_{abc}-Q_{\;\;\;be}^{e}\eta_{ca}-Q_{\;\;\;ce}^{e}\eta_{ba}\right)\label{Gravitational terms for nonmetricity}
\end{eqnarray}
and including sources, the field equation is
\begin{eqnarray}
\mathcal{\mathscr{Q}}_{cba} & = & \frac{4}{\kappa}\frac{\delta\mathcal{L}_{Source}}{\delta Q^{abc}}\label{Field eq for nonmetricity}
\end{eqnarray}
The source term on the right has been called the \emph{hypermomentum}
\cite{HehlvonderHeydeKerlick,HehlMetricAffine,KlemmRavera}. Notice
that we may write $\mathcal{\mathscr{Q}}_{acb}$ as $\mathcal{\mathscr{Q}}_{acb}=-\left(\frac{1}{2}Q_{a\left(bc\right)}-\frac{1}{4}\eta^{de}\left(Q_{b\left(de\right)}\eta_{ca}+Q_{c\left(de\right)}\eta_{ba}\right)\right)$
so that the nonmetric field equation is independent of the nonmetric
2-form $\mathbf{Q}_{a}$ which has been absorbed into the torsion.

We find it helpful to write the variation in the general form
\begin{eqnarray}
\delta S_{Grav} & = & \kappa\int\left(\delta\Omega_{\quad c}^{ab}\mathscr{T}_{\;\;\;ab}^{c}-\frac{1}{4}\delta Q_{\quad a}^{bc}\mathscr{Q}_{\;\;\;bc}^{a}\right)\boldsymbol{\Phi}\label{Gravitational variation}
\end{eqnarray}
where $\Omega_{\quad c}^{ab}$ and $Q_{\quad a}^{bc}$ are the antisymmetric
and symmetric parts of the connection, respectively.

In vacuum, $\mathscr{T}_{\;\;\;bc}^{a}=0$ and $\mathcal{\mathscr{Q}}_{cba}=0$.
The torsion therefore vanishes, while contraction of $\mathcal{\mathscr{Q}}_{cba}=0$
shows that $Q_{\;\;\;ae}^{e}=\frac{1}{4}Q_{\;\;\;ea}^{e}=2W_{a}$
where $W_{b}$ is the Weyl vector. It follows that $Q_{c\left(ab\right)}=2\eta_{c(a}W_{b)}$
so that the full solution for the nonmetricity in vacuum is $Q_{abc}=2\eta_{c(a}W_{b)}+Q_{a\left[bc\right]}$.
The full nonmetricity $\mathbf{Q}_{ab}$ is now given in terms of
$\boldsymbol{\omega}$ and $\mathbf{Q}^{a}$ only, both of which may
be absorbed into a Weyl geometry with torsion \cite{Wheeler2025a}.

\section{Clifford basis and the metric-affine interaction \label{sec:Clifford-basis-and}}

The vacuum Dirac action is
\begin{eqnarray}
S_{D,V} & = & \alpha\int\psi^{\dagger}h\left(i\cancel{\partial}-m\right)\psi\:d^{4}x\label{Dirac action in vacuum}
\end{eqnarray}
where $\psi\in\mathbb{C}^{4}$ is a Dirac spinor, $h$ is the Hermitian
metric $\left\langle \chi,\psi\right\rangle =\chi^{\dagger}h\psi\equiv\bar{\chi}\psi$,
and $\cancel{\partial}=\gamma^{a}\partial_{a}$ where the $\gamma^{a}$
are four $4\times4$ matrices satisfying
\begin{equation}
\left\{ \gamma^{a},\gamma^{b}\right\} =-2\eta^{ab}\mathbf{1}\label{Defining Clifford relation}
\end{equation}
with $a,b,\ldots=0,1,2,3$. The commutators $\sigma^{ab}=\frac{1}{2}\left[\gamma^{a},\gamma^{b}\right]$
span the Lie algebra of the Lorentz group, so that $\Lambda\left(w^{ab}\right)=\exp\left(\frac{1}{2}w_{ab}\sigma^{ab}\right)$
are Lorentz transformations. The action is real, $S_{D}^{*}=S_{D}$,
and the Hermitian metric satisfies $\gamma^{a\dagger}h=h\gamma^{a}$.

In nonflat spacetimes, the Dirac equation requires a covariant derivative.
The connection becomes a Lie-algebra valued 1-form $\boldsymbol{\beta}_{A}G^{A}$
where generators $G^{A}$ form a basis for a spinor representation
of the Lie algebra of the gravitational symmetry. In spaces with compatible
$SO\left(p,q\right)$ invariant metric $\eta_{ab}$, this takes the
form $\frac{1}{2}\boldsymbol{\omega}_{ab}\sigma^{ab}$ where the coefficients
are characterized by antisymmetry $\boldsymbol{\omega}_{ab}=-\boldsymbol{\omega}_{ba}$.
The $\left(p,q\right)$ signature then enters when an index is raised,
$\boldsymbol{\omega}_{\;\;\;b}^{a}=\eta^{ac}\boldsymbol{\omega}_{cb}$.
In particular this applies to a Lorentz connection, precluding any
nonmetricity $\mathbf{Q}_{ab}$ since in this basis nonmetricity depends
entirely on the symmetric part $\mathbf{Q}_{ab}=-\boldsymbol{\omega}_{ab}-\boldsymbol{\omega}_{ba}$.
Therefore, nonmetricity is exactly the Lorentz- or $SO\left(p,q\right)$-violating
part of the spacetime connection.

The requirement for a spinor representation requires generators $G^{A}$
that act linearly on spinors. A complete set of such operators is
given by the Clifford algebra associated with a spin group. We can
accomplish this for a general linear connection by writing a $\mathfrak{gl}\left(4\right)$
basis as elements of either $\mathfrak{Cl}\left(3,1\right)$ or $\mathfrak{Cl}\left(2,2\right)$.

We describe these Clifford algebras here, then identify the resulting
spinor-$\mathfrak{gl}\left(4\right)$ interaction.

\subsection{Cl(3,1)}

The Clifford algebra $\mathfrak{Cl}\left(3,1\right)$ is the quotient
of the free algebra of the $\gamma$-matrices by the symmetric relation
(\ref{Defining Clifford relation}). Equation (\ref{Defining Clifford relation})
allows us to reduce any further symmetric products, so that a basis
for the Clifford algebra is given by all antisymmetric products $\gamma^{a},\gamma^{[a}\gamma^{b]},\gamma^{[a}\gamma^{b}\gamma^{c]},\gamma^{[a}\gamma^{b}\gamma^{c}\gamma^{d]}$.
Appending the identity, these antisymmetric products form a complete
basis. They are generally written in the more convenient form
\begin{eqnarray}
\Gamma^{\Delta} & = & \left\{ 1,\gamma^{a},\sigma^{ab},\gamma_{5}\gamma^{d},\gamma_{5}\right\} \label{Clifford basis}
\end{eqnarray}
where $\ \sigma^{ab}=\frac{1}{2}\left[\gamma^{a},\gamma^{b}\right],\:\gamma_{5}\gamma^{d}=\frac{i}{3!}e_{\;\;\;abc}^{d}\gamma^{a}\gamma^{b}\gamma^{c}$
and $\gamma_{5}=\frac{i}{4!}\varepsilon_{abcd}\gamma^{a}\gamma^{b}\gamma^{c}\gamma^{d}$.
Upper case Greek indices run $\Delta,\Omega,\ldots=1,\ldots,16$.
The 16 matrices $\Gamma^{\Delta}$ span all complex $4\times4$ matrices
and satisfy the orthonormality relation 
\begin{equation}
\frac{1}{4}tr\left(\Gamma^{\Delta}\Gamma^{\Sigma\dagger}\right)=\delta^{\Delta\Sigma}\label{Orthonormality}
\end{equation}
Their completeness is central to our discussion, because it means
we can expand the real, $4\times4$ representation of the \emph{general
linear} Lie algebra as linear combinations $\beta_{\Delta}\Gamma^{\Delta}$.
These linear combinations combine with spinor fields to form source
currents, $\bar{\psi}_{A}\left[\beta_{\Delta}\Gamma^{\Delta}\right]_{\;\;\;B}^{A}\psi^{B}$.

To make the couplings of Dirac fields to nonmetricity explicit, we
use the Dirac representation, $\gamma^{a}=\left(\begin{array}{cc}
1 & 0\\
0 & -1
\end{array}\right)$, and $\gamma^{i}=\left(\begin{array}{cc}
0 & \sigma^{i}\\
-\sigma^{i} & 0
\end{array}\right)$ $\left(i=1,2,3\right)$. In this representation $\left[h\right]_{AB}=\left[\gamma^{0}\right]_{\;\;\;B}^{A}$,
so the Hermitian inner product $\left\langle \chi,\psi\right\rangle =\chi^{\dagger}h\psi\equiv\bar{\chi}\psi$
we have $\bar{\chi}=\chi^{\dagger}h=\chi^{\dagger}\gamma^{0}$ in
the usual way. The Dirac representation puts the remainder of the
Clifford basis $\Gamma^{A}=\left\{ 1,\gamma^{a},\sigma^{ab},\gamma_{5}\gamma^{a},\gamma_{5}\right\} $
in the form 
\begin{eqnarray*}
\begin{array}{ccccccccccccc}
\sigma^{ij} & = & -i\varepsilon_{\;\quad k}^{ij}\left(\begin{array}{cc}
\sigma^{k}\\
 & \sigma^{k}
\end{array}\right) &  &  & \sigma^{0i} & = & \left(\begin{array}{cc}
 & \sigma^{i}\\
\sigma^{i}
\end{array}\right) &  &  & \gamma_{5}\gamma^{0} & = & \left(\begin{array}{cc}
0 & -1\\
1 & 0
\end{array}\right)\\
\gamma_{5}\gamma^{i} & = & \left(\begin{array}{cc}
-\sigma^{i} & 0\\
0 & \sigma^{i}
\end{array}\right) &  &  & \gamma_{5} & = & \left(\begin{array}{cc}
0 & 1\\
1 & 0
\end{array}\right)
\end{array}
\end{eqnarray*}
Here $\sigma^{i}$ are the Pauli matrices and $1$ is the $2\times2$
identity. Manipulation of the gamma matrices is familiar.

While complex linear combinations $\beta_{\Delta}\Gamma^{\Delta}$
can provide the real basis required for $\mathfrak{gl}\left(4\right)$,
it is simpler to make use of the Clifford algebra associated with
$\mathfrak{spin}\left(2,2\right)$ since this admits a real basis
directly. The Clifford algebra $Cl\left(2,2\right)$ is simply related
to $Cl\left(3,1\right)$ making it straightforward to characterize
the contribution of Dirac spinors.

\subsection{Cl(2,2)}

Unlike $\mathfrak{spin}\left(3,1\right)$, the Lie algebra $\mathfrak{spin}\left(2,2\right)$
admits a real representation. This has the advantage of directly providing
a basis for $\mathfrak{gl}\left(4,\mathbb{R}\right)$, while making
only a slight change to the $\mathfrak{spin}\left(3,1\right)$ basis
$\Gamma^{\Delta}$. As above, lower case Latin indices from the beginning
of the alphabet $a,b,\ldots=0,1,2,3,$ refer to either $SO\left(3,1\right)$
or $SO\left(2,2\right)$, while upper case Latin are spinor indices.

A real form of gamma matrices for $\mathfrak{spin}\left(2,2\right)$
be chosen by inserting a single factor of $i$ on $\gamma^{2}$:
\begin{eqnarray*}
\hat{\gamma}^{a} & \equiv & \left(\gamma^{0},\gamma^{1},i\gamma^{2},\gamma^{3}\right)
\end{eqnarray*}
These satisfy $\left\{ \hat{\gamma}^{a},\hat{\gamma}^{b}\right\} =-2\hat{\eta}^{ab}$
where the metric is $\hat{\eta}_{ab}=diag\left(-1,1,-1,1\right)$.
We confirm that $\hat{\sigma}^{AB},\hat{\gamma}_{5}\hat{\gamma}^{A}$
and $\hat{\gamma}_{5}=\gamma_{5}$ are all real, with $\hat{\gamma}_{5}\hat{\gamma}^{A}=\gamma_{5}\hat{\gamma}^{A}$
and
\begin{eqnarray}
\hat{\sigma}^{AB} & = & \frac{1}{2}\left[\hat{\gamma}^{A},\hat{\gamma}^{B}\right]=\left(\sigma^{01},i\sigma^{02},\sigma^{03},i\sigma^{12},i\sigma^{23},\sigma^{31}\right)\label{Cl(2,2) in terms of CL(3,1)}
\end{eqnarray}
Because the matrices $\hat{\Gamma}^{\Delta}=\left\{ 1,\hat{\gamma}^{A},\hat{\sigma}^{AB},\hat{\gamma}_{5}\hat{\gamma}^{A},\hat{\gamma}_{5}\right\} $
are all real, we may write the $\mathfrak{gl}\left(4,\mathbb{R}\right)$
connection as $\boldsymbol{\Sigma}=\mathbf{b}_{\Delta}\hat{\Gamma}^{\Delta}$
with the 1-forms $\mathbf{b}_{\Delta}$ real, $\Delta=1,\ldots,16$.
Because we express $\hat{\Gamma}^{\Delta}$ in terms of $\Gamma^{\Delta}$
we easily find the action on Dirac spinors.

\subsection{The covariant Dirac equation}

Our action is $S=S_{Grav}+S_{D}$ where $S_{Grav}$ is given by Eq.(\ref{Action}).
The Dirac action $S_{D}$ is adapted to metric-affine geometry by
replacing the gradient with a covariant derivative, $\cancel{\partial}=\gamma^{a}e_{a}^{\;\;\;\mu}\partial_{\mu}\rightarrow\cancel{D}=\gamma^{a}e_{a}^{\;\;\;\mu}D_{\mu}$
where
\[
D_{\mu}\psi=\partial_{\mu}\psi-b_{A\mu}\hat{\Gamma}^{A}\psi
\]
With $b_{A\mu}$ real, $b_{A\mu}\hat{\Gamma}^{A}$ is the $GL\left(4\right)$
connection.

The spinor action must be made manifestly real, so we separate the
real vacuum terms \ref{Dirac action in vacuum} from the interaction,
$S_{D}=S_{D,V}-\alpha Re\int b_{aA}\psi^{\dagger}ih\gamma^{a}\hat{\Gamma}^{A}\psi$
to identify the interaction 
\begin{eqnarray}
S_{interaction} & = & -\alpha Re\int b_{aA}\psi^{\dagger}ih\gamma^{a}\hat{\Gamma}^{A}\psi\label{Spinor interaction}
\end{eqnarray}
The contributions to the field equation are found by varying the real
coefficients $b_{aA}$, but in this form it is not clear which terms
will be sources for torsion and which will drive nonmetricity. To
determine this, in the next Section we find the generators of the
$SO\left(3,1\right)$ subgroup of $GL\left(4\right)$ in terms of
the $\left[\hat{\Gamma}\right]_{\;\;\;B}^{A}$.

\section{Nonmetricity vs. torsion: separating the SO(3,1) subgroup in the
Clifford basis \label{sec:Nonmetricity-vs.-torsion:}}

To apply Eqs.(\ref{Field eq for torsion}) and (\ref{Field eq for nonmetricity})
with the Dirac sources of Eq.(\ref{Spinor interaction}) we expand
the connection in the $\mathfrak{spin}\left(2,2\right)$ basis $\mathbf{b}_{A}\hat{\Gamma}^{A}$.
To correctly interpret the results, we need to distinguish the torsion
and nonmetric parts of the connection. Among the $\hat{\Gamma}^{\Delta}$
there must be real combinations that generate the real vector representation
of the Lorentz group $SO\left(3,1\right)$. Varying the $\mathfrak{so}\left(3,1\right)$
generators will give the coupling to torsion, with the remaining independent
combinations giving the field equation for nonmetricity.

As noted in Section \ref{sec:Clifford-basis-and}, when the generators
of any pseudo-orthogonal group $\mathfrak{so}\left(p,q\right)$ are
in doubly covariant form, they are always antisymmetric, $\left[G_{\Delta}\right]_{AB}=-\left[G_{\Delta}\right]_{BA}$.
The signature of the pseudo-orthogonal metric then enters when we
return one index to the raised position, $\left[G_{\Delta}\right]_{\;\;\;B}^{A}=\eta^{AC}\left[G_{\Delta}\right]_{CB}$.
We may therefore identify a basis for $\mathfrak{so}\left(3,1\right)$
by finding the antisymmetric connection forms, $\left[\hat{\Gamma}\right]_{\left[AB\right]}$.
Starting with the real transformations $\left[\hat{\Gamma}\right]_{\;\;\;B}^{A}$,
we may lower $A$ with any convenient nondegenerate real matrix, e.g.,
$h\simeq\gamma^{0}$ or $\hat{h}=i\hat{\gamma}^{0}\hat{\gamma}^{2}$,
since we then take the antisymmetric part and raise with $\eta^{-1}$.
Different choices merely assign different names to the same set of
Lorentz generators. The simplest choice is the diagonal form $h$.

Writing the covariant matrices $\left[h\hat{\Gamma}\right]_{\left[AB\right]}$
explicitly in the Dirac representation, we find the antisymmetric
subset $\left[h\hat{\Gamma}\right]_{\left[AB\right]}\in\left\{ ih\gamma^{2},h\sigma^{01},h\sigma^{03},h\sigma^{31},ih\gamma_{5}\gamma^{2},h\gamma_{5}\right\} $.
Raising indices with $\eta^{AB}\equiv diag\left(-1,1,1,1\right)$
gives a real basis $\hat{\Gamma}^{\Delta_{a}}$ for $\mathfrak{so}\left(3,1\right)$:
\begin{eqnarray*}
 &  & i\eta h\gamma^{2}=\left(\begin{array}{cc}
 & -\sigma^{1}\\
i\sigma^{2}
\end{array}\right),\eta h\hat{\gamma}_{5}\hat{\gamma}^{2}=\left(\begin{array}{cc}
\sigma^{1}\\
 & -i\sigma^{2}
\end{array}\right),\eta h\hat{\gamma}_{5}=\left(\begin{array}{cc}
 & -\sigma^{3}\\
-1
\end{array}\right)\\
 &  & \eta h\sigma^{01}=\left(\begin{array}{cc}
 & -i\sigma^{2}\\
-\sigma^{1}
\end{array}\right),\eta h\sigma^{03}=\left(\begin{array}{cc}
 & -1\\
-\sigma^{3}
\end{array}\right),\eta h\sigma^{31}=\left(\begin{array}{cc}
\sigma^{1}\\
 & i\sigma^{2}
\end{array}\right)
\end{eqnarray*}
It is straightforward to check that these span $\mathfrak{so}\left(3,1\right)$.
In terms of the usual boost and rotation generators, $\left[K^{i}\right]_{\;\;\;b}^{a}=\delta_{0}^{a}\delta_{b}^{i}+\delta^{ia}\delta_{0b}$,
$\left[J^{i}\right]_{\;\;\;b}^{a}=\varepsilon_{\quad b}^{ia}$ respectively
\begin{eqnarray*}
\begin{array}{ccccccc}
K_{x} & = & \frac{1}{2}\left(\eta h\hat{\gamma}_{5}\hat{\gamma}^{2}+\eta h\sigma^{31}\right) &  & J_{x} & = & \frac{1}{2}\left(\eta h\sigma^{01}-i\eta h\gamma^{2}\right)\\
K_{y} & = & -\frac{1}{2}\left(\eta h\hat{\gamma}_{5}+\eta h\sigma^{03}\right) &  & J_{y} & = & \frac{1}{2}\left(\eta h\sigma^{03}-\eta h\hat{\gamma}_{5}\right)\\
K_{z} & = & -\frac{1}{2}\left(i\eta h\gamma^{2}+\eta h\sigma^{01}\right) &  & J_{z} & = & \frac{1}{2}\left(\eta h\sigma^{31}-\eta h\hat{\gamma}_{5}\hat{\gamma}^{2}\right)
\end{array}
\end{eqnarray*}

The remaining 10 combinations
\begin{eqnarray*}
\eta^{AC}\left[h\hat{\Gamma}\right]_{\left(CB\right)} & \in & \hat{\Gamma}^{\Delta_{s}}=\left\{ h1,h\gamma^{0},ih\sigma^{12},ih\sigma^{23},h\gamma_{5}\gamma^{1},h\gamma_{5}\gamma^{3},h\gamma^{1}h\gamma^{3},ih\sigma^{02},h\gamma_{5}\gamma^{0}\right\} 
\end{eqnarray*}
 with $h\hat{\Gamma}$ symmetric, give generators
\begin{eqnarray*}
 &  & \eta h1=\left(\begin{array}{cc}
-\sigma^{3}\\
 & -1
\end{array}\right),\eta h\gamma^{0}=\left(\begin{array}{cc}
-\sigma^{3}\\
 & 1
\end{array}\right),\eta h\gamma^{1}=\left(\begin{array}{cc}
 & -i\sigma^{2}\\
-\sigma^{1}
\end{array}\right),\eta h\gamma^{3}=\left(\begin{array}{cc}
 & -1\\
\sigma^{3}
\end{array}\right),\\
 &  & i\eta h\sigma^{02}=\left(\begin{array}{cc}
 & -\sigma^{1}\\
-i\sigma^{2}
\end{array}\right),i\eta h\sigma^{12}=\left(\begin{array}{cc}
-1\\
 & -\sigma^{3}
\end{array}\right),i\eta h\sigma^{23}=\left(\begin{array}{cc}
-i\sigma^{2}\\
 & -\sigma^{1}
\end{array}\right),\\
 &  & \eta h\hat{\gamma}_{5}\hat{\gamma}^{1}=\left(\begin{array}{cc}
i\sigma^{2}\\
 & -\sigma^{1}
\end{array}\right),\eta h\hat{\gamma}_{5}\hat{\gamma}^{3}=\left(\begin{array}{cc}
1\\
 & -\sigma^{3}
\end{array}\right),\eta h\hat{\gamma}_{5}\hat{\gamma}^{2}=\left(\begin{array}{cc}
\sigma^{1}\\
 & -i\sigma^{2}
\end{array}\right)
\end{eqnarray*}
Varying a real linear combination of these will give our source for
nonmetricity.

The full connection is therefore $\boldsymbol{\Sigma}=\mathbf{b}_{\Delta}\Gamma^{\Delta}=\mathbf{b}_{\Delta_{s}}\Gamma^{\Delta_{s}}+\mathbf{b}_{\Delta_{a}}\Gamma^{\Delta_{a}}$.
Explicitly
\begin{eqnarray}
\Omega_{\;\;\;Bc}^{A}=b_{c\Delta_{a}}\left[\Gamma^{\Delta_{a}}\right]_{\;\;\;B}^{A} & = & \left[ib_{c2}\eta h\gamma^{2}+b_{c01}\eta h\sigma^{01}+b_{c03}\eta h\sigma^{03}+b_{c31}\eta h\sigma^{31}+ib_{c52}\eta h\gamma_{5}\gamma^{2}+b_{c5}\eta h\gamma_{5}\right]_{\;\;\;B}^{A}\nonumber \\
Q_{\;\;\;Bc}^{A}=b_{c\Delta_{s}}\left[\Gamma^{\Delta_{s}}\right]_{\;\;\;B}^{A} & = & \left[b_{c}\eta h1\right.+b_{c0}\eta h\gamma^{0}+b_{c1}\eta h\gamma^{1}+b_{c3}\eta h\gamma^{3}+ib_{c02}\eta h\sigma^{02}+ib_{c12}\eta h\sigma^{12}+ib_{c23}\eta h\sigma^{23}\nonumber \\
 &  & +b_{50}\eta h\gamma_{5}\gamma^{0}+b_{c51}\eta h\gamma_{5}\gamma^{1}+\left.b_{c53}\eta h\gamma_{5}\gamma^{3}\right]_{\;\;\;B}^{A}\label{Explicit connection}
\end{eqnarray}
Varying $\left(b_{c\mathfrak{ij}},b_{c2},b_{c52},b_{c5}\right)_{\mathfrak{i},\mathfrak{j}=0,1,3}$
will give the source for torsion while varying $\left(b_{c},b_{c\mathfrak{i}},b_{c2\mathfrak{i}},b_{c5\mathfrak{i}}\right)_{\mathfrak{i}=0,1,3}$
will give the source for nonmetricity.

\section{The field equations \label{sec:The-field-equations}}

Writing the gravitational variation (\ref{Gravitational variation})
in the $\mathfrak{Cl}\left(2,2\right)$ expansion, we use the antisymmetry
of $\left[\Gamma^{\Delta_{a}}\right]^{AB}$ and the symmetry of $\left[\Gamma^{\Delta_{s}}\right]^{AB}$
to write the products as traces.
\begin{eqnarray*}
\delta\Omega_{\qquad c}^{ab}\mathscr{T}_{\;\;\;ab}^{c} & = & b_{c\Delta_{a}}\left[\Gamma^{\Delta_{a}}\right]^{AB}\mathscr{T}_{\;\;\;AB}^{c}=-b_{c\Delta_{a}}tr\left(\Gamma^{\Delta_{a}}\mathscr{T}^{c}\right)\\
-\frac{1}{4}\delta Q_{\quad a}^{bc}\mathscr{Q}_{\;\;\;bc}^{a} & = & -\frac{1}{4}b_{c\Delta_{s}}\left[\Gamma^{\Delta_{s}}\right]^{AB}\mathscr{Q}_{\;\;\;AB}^{c}=-\frac{1}{4}b_{c\Delta_{s}}tr\left(\Gamma^{\Delta_{s}}\mathscr{Q}^{c}\right)
\end{eqnarray*}
where $\mathscr{T}_{\;\;\;ab}^{c}$ and $\mathscr{Q}_{\;\;\;bc}^{a}$
are given by Eqs.(\ref{ECSK source for torsion}) and (\ref{Gravitational terms for nonmetricity}),
respectively. The field equations follow from the variation $\delta S_{Grav}+\delta S_{D}=0$.
\begin{eqnarray}
-\kappa\delta b_{c\Delta_{a}}tr\left(\Gamma^{\Delta_{a}}\mathscr{T}^{c}\right)-\frac{\kappa}{4}\delta b_{c\Delta_{s}}tr\left(\Gamma^{\Delta_{s}}\mathscr{Q}_{c}\right) & = & \alpha Re\left(i\psi^{\dagger}h\gamma^{c}\delta b_{\Delta c}\hat{\Gamma}^{\Delta}\psi\right)\label{Field equation}
\end{eqnarray}
Next, we expand each side explicitly.

\subsection{Gravitational interaction}

Substituting the explicit connection from Eqs.(\ref{Explicit connection})
into Eq.(\ref{Gravitational interaction}), we find the gravitational
contribution to the field equation by expanding the sums on $\Delta_{a}$
and $\Delta_{s}$. Notice that once we take the traces, the expressions
are real components of Lorentzian matrices, so we revert to lower
case Latin indices. Writing out the full sum,
\begin{eqnarray}
\delta\mathcal{L}_{Grav} & = & \kappa\delta\Omega_{\quad c}^{AB}\mathcal{T}_{\;\;\;AB}^{c}-\frac{\kappa}{4}\delta Q^{ABc}\mathscr{Q}_{c\left(AB\right)}\nonumber \\
 & = & -\kappa\delta b_{c2}tr\left[i\eta h\gamma^{2}\mathcal{\mathscr{T}}^{c}\right]-\kappa\delta b_{c01}tr\left[\eta h\sigma^{01}\mathcal{\mathscr{T}}^{c}\right]-\kappa\delta b_{c03}tr\left[\eta h\sigma^{03}\mathcal{\mathscr{T}}^{c}\right]\nonumber \\
 &  & -\kappa\delta b_{c13}tr\left[\eta h\sigma^{13}\mathcal{\mathscr{T}}^{c}\right]-\kappa\delta b_{c52}tr\left[i\eta h\gamma_{5}\gamma^{2}\mathcal{\mathscr{T}}^{c}\right]-\kappa\delta b_{c5}tr\left[\eta h\gamma_{5}\mathcal{\mathscr{T}}^{c}\right]\nonumber \\
 &  & -\frac{\kappa}{4}\delta b_{c}tr\left[\eta h1\mathcal{\mathscr{Q}}^{c}\right]-\frac{\kappa}{4}\delta b_{c0}tr\left[\eta h\gamma^{0}\mathcal{\mathscr{Q}}^{c}\right]-\frac{\kappa}{4}\delta b_{c1}tr\left[\eta h\gamma^{1}\mathcal{\mathscr{Q}}^{c}\right]-\frac{\kappa}{4}\delta b_{c3}tr\left[\eta h\gamma^{3}\mathcal{\mathscr{Q}}^{c}\right]\nonumber \\
 &  & -\frac{\kappa}{4}\delta b_{c02}tr\left[i\eta h\sigma^{02}\mathcal{\mathscr{Q}}^{c}\right]-\frac{\kappa}{4}\delta b_{c12}tr\left[i\eta h\sigma^{12}\mathcal{\mathscr{Q}}^{c}\right]-\frac{\kappa}{4}\delta b_{c23}tr\left[i\eta h\sigma^{23}\mathcal{\mathscr{Q}}^{c}\right]\nonumber \\
 &  & -\frac{\kappa}{4}\delta b_{c50}tr\left[\eta h\gamma_{5}\gamma^{0}\mathcal{\mathscr{Q}}^{c}\right]-\frac{\kappa}{4}\delta b_{c51}tr\left[\eta h\gamma_{5}\gamma^{1}\mathcal{\mathscr{Q}}^{c}\right]-\frac{\kappa}{4}\delta b_{c53}tr\left[\eta h\gamma_{5}\gamma^{3}\mathcal{\mathscr{Q}}^{c}\right]\label{Gravitational interaction}
\end{eqnarray}
we then compute each of the 16 traces. The result is a collection
of linear combinations of components of $\mathcal{T}_{\;\;\;ab}^{c}$.
We find the explicit combinations by carrying out the traces in the
Dirac representation.

For the torsion, $\mathcal{T}_{\quad ab}^{c}$ these we find
\begin{eqnarray}
-\left[i\eta h\gamma^{2}\right]_{\;\;\;b}^{a}\mathcal{T}_{\quad a}^{cb} & = & \mathcal{T}_{\quad30}^{c}+\mathcal{T}_{\quad21}^{c}-\mathcal{T}_{\quad12}^{c}-\mathcal{T}_{\quad03}^{c}\nonumber \\
-\left[\eta h\sigma^{01}\right]_{\;\;\;b}^{a}\mathcal{T}_{\quad a}^{cb} & = & \mathcal{T}_{\quad30}^{c}-\mathcal{T}_{\quad21}^{c}+\mathcal{T}_{\quad12}^{c}-\mathcal{T}_{\quad03}^{c}\nonumber \\
-\left[i\eta h\gamma_{5}\gamma^{2}\right]_{\;\;\;b}^{a}\mathcal{T}_{\quad a}^{cb} & = & -\mathcal{T}_{\quad10}^{c}+\mathcal{T}_{\quad01}^{c}+\mathcal{T}_{\quad32}^{c}-\mathcal{T}_{\quad23}^{c}\nonumber \\
-\left[\eta h\sigma^{31}\right]_{\;\;\;b}^{a}\mathcal{T}_{\quad a}^{cb} & = & -\mathcal{T}_{\quad10}^{c}+\mathcal{T}_{\quad01}^{c}-\mathcal{T}_{\quad32}^{c}+\mathcal{T}_{\quad23}^{c}\nonumber \\
-\left[\eta h\gamma_{5}\right]_{\;\;\;b}^{a}\mathcal{T}_{\quad a}^{cb} & = & \mathcal{T}_{\quad20}^{c}-\mathcal{T}_{\quad31}^{c}-\mathcal{T}_{\quad02}^{c}+\mathcal{T}_{\quad13}^{c}\nonumber \\
-\left[\eta h\sigma^{03}\right]_{\;\;\;b}^{a}\mathcal{T}_{\quad a}^{cb} & = & \mathcal{T}_{\quad20}^{c}+\mathcal{T}_{\quad31}^{c}-\mathcal{T}_{\quad02}^{c}-\mathcal{T}_{\quad13}^{c}\label{Lgrav Torsion}
\end{eqnarray}
The traces for nonmetricity are
\begin{eqnarray}
tr\left[\eta h1\mathcal{\mathscr{Q}}^{c}\right] & = & \mathcal{\mathscr{Q}}_{\quad00}^{c}+\mathcal{\mathscr{Q}}_{\quad11}^{c}-\mathcal{\mathscr{Q}}_{\quad22}^{c}-\mathcal{\mathscr{Q}}_{\quad33}^{c}\nonumber \\
tr\left[\eta h\gamma^{0}\mathcal{\mathscr{Q}}^{c}\right] & = & \mathcal{\mathscr{Q}}_{\quad00}^{c}+\mathcal{\mathscr{Q}}_{\quad11}^{c}+\mathcal{\mathscr{Q}}_{\quad22}^{c}+\mathcal{\mathscr{Q}}_{\quad33}^{c}\nonumber \\
tr\left[\eta h\gamma^{1}\mathcal{\mathscr{Q}}^{c}\right] & = & -\mathcal{\mathscr{Q}}_{\quad30}^{c}+\mathcal{\mathscr{Q}}_{\quad21}^{c}+\mathcal{\mathscr{Q}}_{\quad12}^{c}-\mathcal{\mathscr{Q}}_{\quad03}^{c}\nonumber \\
tr\left[\eta h\gamma^{3}\mathcal{\mathscr{Q}}^{c}\right] & = & -\mathcal{\mathcal{\mathscr{Q}}}_{\quad20}^{c}-\mathcal{\mathscr{Q}}_{\quad31}^{c}-\mathcal{\mathscr{Q}}_{\quad02}^{c}-\mathcal{\mathscr{Q}}_{\quad13}^{c}\nonumber \\
tr\left[i\eta h\sigma^{02}\mathcal{\mathscr{Q}}^{c}\right] & = & -\mathcal{\mathscr{Q}}_{\quad30}^{c}-\mathcal{\mathscr{Q}}_{\quad12}^{c}-\mathcal{\mathscr{Q}}_{\quad03}^{c}-\mathcal{\mathscr{Q}}_{\quad21}^{c}\nonumber \\
tr\left[i\eta h\sigma^{12}\mathcal{\mathscr{Q}}^{c}\right] & = & \mathcal{\mathscr{Q}}_{\quad00}^{c}-\mathcal{\mathscr{Q}}_{\quad11}^{c}-\mathcal{\mathscr{Q}}_{\quad22}^{c}+\mathcal{\mathscr{Q}}_{\quad33}^{c}\nonumber \\
tr\left[i\eta h\sigma^{32}\mathcal{\mathscr{Q}}^{c}\right] & = & \mathcal{\mathscr{Q}}_{\quad10}^{c}+\mathcal{\mathscr{Q}}_{\quad01}^{c}+\mathcal{\mathscr{Q}}_{\quad32}^{c}+\mathcal{\mathscr{Q}}_{\quad23}^{c}\nonumber \\
tr\left[\eta h\gamma_{5}\gamma^{0}\mathcal{\mathscr{Q}}^{c}\right] & = & \mathcal{\mathcal{\mathscr{Q}}}_{\quad20}^{c}-\mathcal{\mathscr{Q}}_{\quad31}^{c}+\mathcal{\mathscr{Q}}_{\quad02}^{c}-\mathcal{\mathscr{Q}}_{\quad13}^{c}\nonumber \\
tr\left[\eta h\gamma_{5}\gamma^{1}\mathcal{\mathscr{Q}}^{c}\right] & = & \mathcal{\mathscr{Q}}_{\quad10}^{c}+\mathcal{\mathscr{Q}}_{\quad01}^{c}-\mathcal{\mathscr{Q}}_{\quad32}^{c}-\mathcal{\mathscr{Q}}_{\quad23}^{c}\nonumber \\
tr\left[\eta h\gamma_{5}\gamma^{3}\mathcal{\mathscr{Q}}^{c}\right] & = & -\mathcal{\mathscr{Q}}_{\quad00}^{c}+\mathcal{\mathscr{Q}}_{\quad11}^{c}-\mathcal{\mathscr{Q}}_{\quad22}^{c}+\mathcal{\mathscr{Q}}_{\quad33}^{c}\label{Lgrav Nonmetric}
\end{eqnarray}
For each $c=0,1,2,3$ these are equated to scalars built from Dirac
spinors.

\subsection{Spinor interaction}

The sum over $\Delta$ in the spinor contribution to Eq.(\ref{Field equation})
gives
\begin{eqnarray}
\delta\mathcal{L}_{interaction} & = & -\alpha Re\left(\psi^{\dagger}\left(i\delta b_{a01}h\gamma^{a}\eta h\sigma^{01}+i\delta b_{a03}h\gamma^{a}\eta h\sigma^{03}+i\delta b_{a13}h\gamma^{a}\eta h\sigma^{13}\right)\psi\right)\nonumber \\
 &  & -\alpha Re\left(\psi^{\dagger}\left(i\delta b_{a2}ih\gamma^{a}\eta h\gamma^{2}+i\delta b_{a52}ih\gamma^{a}\eta h\gamma_{5}\gamma^{2}+i\delta b_{a5}h\gamma^{a}\eta h\gamma_{5}\right)\psi\right)\nonumber \\
 &  & -\alpha\delta b_{c}Re\psi^{\dagger}\left(ih\gamma^{c}\eta h1\right)\psi-\alpha Re\left(\psi^{\dagger}\left(i\delta b_{a0}h\gamma^{a}\eta h\gamma^{0}+i\delta b_{a1}h\gamma^{a}\eta h\gamma^{1}+i\delta b_{a3}h\gamma^{a}\eta h\gamma^{3}\right)\psi\right)\nonumber \\
 &  & -\alpha\left(Re\psi^{\dagger}\left(i\delta b_{a50}h\gamma^{a}\eta h\gamma_{5}\gamma^{0}+i\delta b_{a51}h\gamma^{a}\eta h\gamma_{5}\gamma^{1}+i\delta b_{a53}h\gamma^{a}\eta h\gamma_{5}\gamma^{3}\right)\psi\right)\nonumber \\
 &  & -\alpha Re\left(\psi^{\dagger}\left(i\delta b_{a02}ih\gamma^{a}\eta h\sigma^{02}+i\delta b_{a12}ih\gamma^{a}\eta h\sigma^{12}+i\delta b_{a32}ih\gamma^{a}\eta h\sigma^{32}\right)\psi\right)\label{Expanded spinor interaction}
\end{eqnarray}
where there remains a sum on $a=0,1,2,3$. We list the torsion coefficients
$\left\{ b_{a01},b_{a03},b_{a13},b_{a2},b_{a52},b_{a5}\right\} $
first.

The breaking of Lorentz symmetry by the general linear group disrupts
the usual systematic index notation so that each term must be computed
individually. The computation of each term is straightforward. Throughout
we choose the spinor components as
\begin{eqnarray*}
\psi & = & \left(\begin{array}{c}
\mu\\
\nu\\
\rho\\
\sigma
\end{array}\right),\psi^{\dagger}=\left(\bar{\mu},\bar{\nu},\bar{\rho},\bar{\sigma}\right)
\end{eqnarray*}
where the over bar denotes complex conjugation. For example, the third
term in Eq.(\ref{Expanded spinor interaction}), $-\alpha b_{31}Re\left(\psi^{\dagger}ih\gamma^{3}\eta h\gamma^{1}\psi\right)$
becomes
\begin{eqnarray*}
-\alpha Re\left(\psi^{\dagger}b_{31}ih\gamma^{3}\eta h\gamma^{1}\psi\right) & = & -\alpha b_{31}Re\left(i\psi^{\dagger}\left(\begin{array}{cc}
1 & 0\\
0 & -1
\end{array}\right)\left(\begin{array}{cc}
 & \sigma^{3}\\
-\sigma^{3}
\end{array}\right)\left(\begin{array}{cc}
-\sigma^{3}\\
 & 1
\end{array}\right)\left(\begin{array}{cc}
1\\
 & -1
\end{array}\right)\left(\begin{array}{cc}
 & \sigma^{1}\\
-\sigma^{1}
\end{array}\right)\psi\right)\\
 & = & -i\alpha b_{31}\left(\bar{\mu}\nu-\bar{\nu}\mu\right)
\end{eqnarray*}
This simplification is carried out for each of the 64 terms in Eq.(\ref{Expanded spinor interaction}).
The resulting Dirac scalars are given in the Appendix \ref{sec:Appendix-Spinor couplings}.

\subsection{Field equations}

This Subsection contains our principal results.
\begin{flushleft}
The field equations $\delta\mathcal{L}_{Grav}=-\delta\mathcal{L}_{Interaction}$
follow by equating the variation coefficients $\delta b_{cA}$. For
the torsion these become the 24 equations
\begin{eqnarray*}
\left[\eta hi\gamma^{2}\right]_{\;\;\;b}^{a}\mathcal{T}_{\quad a}^{cb} & = & -\frac{\alpha}{\kappa}Re\left(\psi^{\dagger}ih\gamma^{c}i\eta h\gamma^{2}\psi\right)\\
\left[\eta h\sigma^{01}\right]_{\;\;\;b}^{a}\mathcal{T}_{\quad a}^{cb} & = & -\frac{\alpha}{\kappa}Re\left(\psi^{\dagger}ih\gamma^{c}\eta h\sigma^{01}\psi\right)\\
\left[\eta h\gamma_{5}i\gamma^{2}\right]_{\;\;\;b}^{a}\mathcal{T}_{\quad a}^{cb} & = & -\frac{\alpha}{\kappa}Re\left(\psi^{\dagger}ih\gamma^{c}i\eta h\gamma_{5}\gamma^{2}\psi\right)\\
\left[\eta h\sigma^{31}\right]_{\;\;\;b}^{a}\mathcal{T}_{\quad a}^{cb} & = & -\frac{\alpha}{\kappa}Re\left(\psi^{\dagger}ih\gamma^{c}\eta h\sigma^{31}\psi\right)\\
\left[\eta h\gamma_{5}\right]_{\;\;\;b}^{a}\mathcal{T}_{\quad a}^{cb} & = & -\frac{\alpha}{\kappa}Re\left(\psi^{\dagger}ih\gamma^{c}\eta h\gamma_{5}\psi\right)\\
\left[\eta h\sigma^{03}\right]_{\;\;\;b}^{a}\mathcal{T}_{\quad a}^{cb} & = & -\frac{\alpha}{\kappa}Re\left(\psi^{\dagger}ih\gamma^{c}\eta h\sigma^{03}\psi\right)
\end{eqnarray*}
This is the $\mathfrak{spin}\left(2,2\right)$ decomposition of the
torsion, with each independent projection sourced by a different current.
The $\mathfrak{spin}\left(2,2\right)$ decomposition of the nonmetricity
is comprised of the remaining 40 equations.
\begin{eqnarray*}
\left[\eta h\right]_{\;\;\;b}^{c}\mathscr{Q}_{\quad c}^{ab} & = & -\frac{4\alpha}{\kappa}Re\left(\psi^{\dagger}ih\gamma^{a}\eta h\psi\right)\\
\left[\eta h\gamma^{0}\right]_{\;\;\;b}^{c}\mathscr{Q}_{\quad c}^{ab} & = & -\frac{4\alpha}{\kappa}Re\left(\psi^{\dagger}ih\gamma^{a}\eta h\gamma^{0}\psi\right)\\
\left[\eta h\gamma^{1}\right]_{\;\;\;b}^{c}\mathscr{Q}_{\quad c}^{ab} & = & -\frac{4\alpha}{\kappa}Re\left(\psi^{\dagger}ih\gamma^{a}\eta h\gamma^{1}\psi\right)\\
\left[\eta h\gamma^{3}\right]_{\;\;\;b}^{c}\mathscr{Q}_{\quad c}^{ab} & = & -\frac{4\alpha}{\kappa}Re\left(\psi^{\dagger}ih\gamma^{a}\eta h\gamma^{3}\psi\right)\\
\left[\eta h\gamma_{5}\gamma^{0}\right]_{\;\;\;b}^{c}\mathscr{Q}_{\quad c}^{ab} & = & -\frac{4\alpha}{\kappa}Re\left(\psi^{\dagger}ih\gamma^{a}\eta h\gamma_{5}\gamma^{0}\psi\right)\\
\left[\eta h\gamma_{5}\gamma^{1}\right]_{\;\;\;b}^{c}\mathscr{Q}_{\quad c}^{ab} & = & -\frac{4\alpha}{\kappa}Re\left(\psi^{\dagger}ih\gamma^{a}\eta h\gamma_{5}\gamma^{1}\psi\right)\\
\left[\eta h\gamma_{5}\gamma^{3}\right]_{\;\;\;b}^{c}\mathscr{Q}_{\quad c}^{ab} & = & -\frac{4\alpha}{\kappa}Re\left(\psi^{\dagger}ih\gamma^{a}\eta h\gamma_{5}\gamma^{3}\psi\right)\\
\left[i\eta h\sigma^{02}\right]_{\;\;\;b}^{c}\mathscr{Q}_{\quad c}^{ab} & = & -\frac{4\alpha}{\kappa}Re\left(\psi^{\dagger}ih\gamma^{a}i\eta h\sigma^{02}\psi\right)\\
\left[i\eta h\sigma^{12}\right]_{\;\;\;b}^{c}\mathscr{Q}_{\quad c}^{ab} & = & -\frac{4\alpha}{\kappa}Re\left(\psi^{\dagger}ih\gamma^{a}i\eta h\sigma^{12}\psi\right)\\
\left[i\eta h\sigma^{23}\right]_{\;\;\;b}^{c}\mathscr{Q}_{\quad c}^{ab} & = & -\frac{4\alpha}{\kappa}Re\left(\psi^{\dagger}ih\gamma^{a}i\eta h\sigma^{23}\psi\right)
\end{eqnarray*}
\par\end{flushleft}

The sources on the right are built from the components of $\psi$,
but because of the factor of $\eta$ they do not correspond directly
to the usual Dirac currents. To see the detailed form of the sources,
we replace each \emph{component} on the right with their scalar expansions.
These are listed in the Appendix \ref{sec:Appendix-Spinor couplings}.
We replace the corresponding linear combinations on the left with
the expansions in Eqs.(\ref{Lgrav Torsion}) and (\ref{Lgrav Nonmetric}),
and it is then easy to solve for the individual components $\mathcal{T}_{\;\;\;bc}^{a}$
of the torsion and $\mathscr{Q}_{\;\;\;bc}^{a}$ of the nonmetricity.

Write the real, imaginary, and diagonal parts of complex products
as the real numbers:
\begin{eqnarray*}
R_{\bar{\alpha}\beta} & = & \frac{1}{2}\left(\bar{\alpha}\beta+\bar{\beta}\alpha\right)\\
I_{\bar{\alpha}\beta} & = & \frac{1}{2i}\left(\bar{\alpha}\beta-\bar{\beta}\alpha\right)\\
D_{\bar{\alpha}\alpha} & = & \bar{\alpha}\alpha
\end{eqnarray*}
Then the results for torsion are:
\begin{eqnarray}
\mathcal{T}_{\;\;\;ab}^{0} & = & \frac{\alpha}{\kappa}\left(\begin{array}{cccc}
0 & 0 & 0 & 0\\
0 & 0 & -I_{\bar{\nu}\rho} & -I_{\bar{\nu}\sigma}\\
0 & I_{\bar{\nu}\rho} & 0 & -I_{\bar{\rho}\sigma}\\
0 & I_{\bar{\nu}\sigma} & I_{\bar{\rho}\sigma} & 0
\end{array}\right)\nonumber \\
\nonumber \\\nonumber \\\mathcal{T}_{\;\;\;ab}^{1} & = & \frac{\alpha}{2\kappa}\left(\begin{array}{cccc}
0 & I_{\bar{\nu}\sigma}+I_{\bar{\mu}\rho} & I_{\bar{\rho}\sigma}+I_{\bar{\mu}\nu} & 0\\
-I_{\bar{\nu}\sigma}-I_{\bar{\mu}\rho} & 0 & 0 & I_{\bar{\mu}\nu}-I_{\bar{\rho}\sigma}\\
-I_{\bar{\rho}\sigma}-I_{\bar{\mu}\nu} & 0 & 0 & I_{\bar{\mu}\rho}-I_{\bar{\nu}\sigma}\\
0 & I_{\bar{\rho}\sigma}-I_{\bar{\mu}\nu} & I_{\bar{\nu}\sigma}-I_{\bar{\mu}\rho} & 0
\end{array}\right)\nonumber \\
\nonumber \\\nonumber \\\mathcal{T}_{\;\;\;ab}^{2} & = & \frac{\alpha}{2\kappa}\left(\begin{array}{cccc}
0 & R_{\bar{\mu}\rho}-R_{\bar{\nu}\sigma} & -R_{\bar{\mu}\nu}-R_{\bar{\rho}\sigma} & \left(D_{\bar{\mu}\mu}-D_{\bar{\sigma}\sigma}\right)\\
R_{\bar{\nu}\sigma}-R_{\bar{\mu}\rho} & 0 & \left(D_{\bar{\nu}\nu}+D_{\bar{\rho}\rho}\right) & R_{\bar{\rho}\sigma}-R_{\bar{\mu}\nu}\\
R_{\bar{\mu}\nu}+R_{\bar{\rho}\sigma} & -\left(D_{\bar{\nu}\nu}+D_{\bar{\rho}\rho}\right) & 0 & -\left(R_{\bar{\mu}\rho}+R_{\bar{\nu}\sigma}\right)\\
-\left(D_{\bar{\mu}\mu}-D_{\bar{\sigma}\sigma}\right) & R_{\bar{\mu}\nu}-R_{\bar{\rho}\sigma} & \left(R_{\bar{\mu}\rho}+R_{\bar{\nu}\sigma}\right) & 0
\end{array}\right)\nonumber \\
\nonumber \\\nonumber \\\mathcal{T}_{\;\;\;ab}^{3} & = & \frac{\alpha}{2\kappa}\left(\begin{array}{cccc}
0 & I_{\bar{\nu}\rho}-I_{\bar{\mu}\sigma} & 0 & -\left(I_{\bar{\rho}\sigma}+I_{\bar{\mu}\nu}\right)\\
I_{\bar{\mu}\sigma}-I_{\bar{\nu}\rho} & 0 & I_{\bar{\mu}\nu}-I_{\bar{\rho}\sigma} & 0\\
0 & I_{\bar{\rho}\sigma}-I_{\bar{\mu}\nu} & 0 & -\left(I_{\bar{\mu}\sigma}+I_{\bar{\nu}\rho}\right)\\
I_{\bar{\rho}\sigma}+I_{\bar{\mu}\nu} & 0 & I_{\bar{\mu}\sigma}+I_{\bar{\nu}\rho} & 0
\end{array}\right)\label{Torsion solution}
\end{eqnarray}

The results for nonmetricity are:
\begin{eqnarray}
\mathscr{Q}_{\quad ab}^{0} & = & \frac{4\alpha}{\kappa}\left(\begin{array}{cccc}
0 & I_{\bar{\mu}\nu} & I_{\bar{\mu}\rho} & I_{\bar{\mu}\sigma}\\
I_{\bar{\mu}\nu} & 0 & 0 & 0\\
I_{\bar{\mu}\rho} & 0 & 0 & 0\\
I_{\bar{\mu}\sigma} & 0 & 0 & 0
\end{array}\right)\nonumber \\
\mathscr{Q}_{\quad ab}^{1} & = & \frac{2\alpha}{\kappa}\left(\begin{array}{cccc}
2I_{\bar{\mu}\sigma} & I_{\bar{\mu}\rho}-I_{\bar{\nu}\sigma} & I_{\bar{\mu}\nu}-I_{\bar{\rho}\sigma} & 0\\
I_{\bar{\mu}\rho}-I_{\bar{\nu}\sigma} & -2I_{\bar{\nu}\rho} & 0 & I_{\bar{\mu}\nu}+I_{\bar{\rho}\sigma}\\
I_{\bar{\mu}\nu}-I_{\bar{\rho}\sigma} & 0 & 2I_{\bar{\nu}\rho} & I_{\bar{\mu}\rho}+I_{\bar{\nu}\sigma}\\
0 & I_{\bar{\mu}\nu}+I_{\bar{\rho}\sigma} & I_{\bar{\mu}\rho}+I_{\bar{\nu}\sigma} & 2I_{\bar{\mu}\sigma}
\end{array}\right)\nonumber \\
\mathscr{Q}_{\quad ab}^{2} & = & \frac{2\alpha}{\kappa}\left(\begin{array}{cccc}
-2R_{\bar{\mu}\sigma} & R_{\bar{\nu}\sigma}+R_{\bar{\mu}\rho} & R_{\bar{\rho}\sigma}-R_{\bar{\mu}\nu} & 2\left(D_{\bar{\mu}\mu}+D_{\bar{\sigma}\sigma}\right)\\
R_{\bar{\nu}\sigma}+R_{\bar{\mu}\rho} & -2R_{\bar{\nu}\rho} & 2\left(D_{\bar{\nu}\nu}-D_{\bar{\rho}\rho}\right) & -R_{\bar{\mu}\nu}-R_{\bar{\rho}\sigma}\\
R_{\bar{\rho}\sigma}-R_{\bar{\mu}\nu} & 2\left(D_{\bar{\nu}\nu}-D_{\bar{\rho}\rho}\right) & 2R_{\bar{\nu}\rho} & R_{\bar{\nu}\sigma}-R_{\bar{\mu}\rho}\\
2\left(D_{\bar{\mu}\mu}+D_{\bar{\sigma}\sigma}\right) & -R_{\bar{\mu}\nu}-R_{\bar{\rho}\sigma} & R_{\bar{\nu}\sigma}-R_{\bar{\mu}\rho} & -2R_{\bar{\mu}\sigma}
\end{array}\right)\nonumber \\
\mathscr{Q}_{\quad ab}^{3} & = & -\frac{2\alpha}{\kappa}\left(\begin{array}{cccc}
-2I_{\bar{\mu}\rho} & I_{\bar{\nu}\rho}+I_{\bar{\mu}\sigma} & 0 & I_{\bar{\rho}\sigma}+I_{\bar{\mu}\nu}\\
I_{\bar{\nu}\rho}+I_{\bar{\mu}\sigma} & -2I_{\bar{\nu}\sigma} & I_{\bar{\rho}\sigma}-I_{\bar{\mu}\nu} & 0\\
0 & I_{\bar{\rho}\sigma}-I_{\bar{\mu}\nu} & -2I_{\bar{\mu}\rho} & I_{\bar{\nu}\rho}-I_{\bar{\mu}\sigma}\\
I_{\bar{\rho}\sigma}+I_{\bar{\mu}\nu} & 0 & I_{\bar{\nu}\rho}-I_{\bar{\mu}\sigma} & 2I_{\bar{\nu}\sigma}
\end{array}\right)\label{Nonmetricty solution}
\end{eqnarray}
Note that the nonmetricity has vanishing trace, $\eta^{ab}\mathscr{Q}_{\quad ab}^{c}=0$,
showing that the Dirac equation does not couple to the Weyl vector,
in agreement with \cite{HochbergPlunien}.

\section{Summary and special cases}

\selectlanguage{american}%
We studied the coupling of metric-affine gravity as a $GL\left(4\right)$
gauge theory, with a Dirac spinor field. The principal difficulty
to overcome is that $GL\left(4\right)$ has no natural spinor representation
and does not preserve the Lorentz metric required to define one. The
connection, however, is Lie algebra valued, and the Lie algebra $\mathfrak{gl}\left(4\right)$
is isomorphic to Clifford algebra $Cl(3,1)$. Both are isomorphic
to the real form of the Clifford algebra $Cl\left(2,2\right)$, so
we expanded the $\mathfrak{gl}\left(4\right)$-valued connection as
a real linear combination of the $Cl\left(2,2\right)$ basis. Replacing
$\tilde{\gamma}^{2}\rightarrow i\gamma^{2}$ where $\tilde{\gamma}^{2}$
lies in the $Cl\left(2,2\right)$ basis and $\gamma^{2}$ lies in
the $Cl\left(3,1\right)$ basis then gave the expansion of the $\mathfrak{gl}\left(4\right)$
connection in terms of Dirac matrices. This allows coupling to the
Dirac field.

To separate contributions to torsion from sources for nonmetricity
we identified the $\mathfrak{so}\left(3,1\right)$ subalgebra within
$\mathfrak{gl}\left(4\right)$. That subset of generators was identified
with couplings to torsion, with the remainder of the connection coupling
to nonmetricity.

\selectlanguage{english}%
To carry out the variation, we wrote the $GL\left(4\right)$ curvature,
separating torsion and nonmetric dependence from the usual Riemannian
curvature. Variation showed the usual vanishing of torsion in vacuum,
while the nonmetricity reduced to those parts expressible as a Weyl
geometry with torsion. Similarly, we separated the spinor action into
vacuum, torsion, and nonmetricity contributions.

\selectlanguage{american}%
Variation leads to Eq.(\ref{Field equation}),
\begin{eqnarray*}
-\kappa\delta b_{c\Delta_{a}}tr\left(\Gamma^{\Delta_{a}}\mathscr{T}^{c}\right)-\frac{\kappa}{4}\delta b_{c\Delta_{s}}tr\left(\Gamma^{\Delta_{s}}\mathscr{Q}_{c}\right) & = & \alpha Re\left(i\psi^{\dagger}h\gamma^{c}\delta b_{\Delta c}\hat{\Gamma}^{\Delta}\psi\right)
\end{eqnarray*}
where $tr\left(\Gamma^{\Delta_{a}}\mathscr{T}^{c}\right)$ and $tr\left(\Gamma^{\Delta_{s}}\mathscr{Q}_{c}\right)$
are traces over the gamma matrix expansions. These give linear combinations
of certain torsion and nonmetricity components. The right side gives
the corresponding combinations of Dirac currents.

\selectlanguage{english}%
Our main results, the Dirac sources for torsion and nonmetricity in
metric-affine gravity, are presented as explicit matrices built from
spinor components in Eqs.(\ref{Torsion solution}) and (\ref{Nonmetricty solution}).
This specificity is necessary because the usual covariant notation
breaks down in many expressions. The 64 real scalars constructed from
Dirac spinors are collected in the Appendix.

To conclude, we gain some insight by looking at restricted spinors.
For a spin-up electron at rest, $\psi$ reduces to $\psi_{e0}=\left(\mu,0,0,0\right)$.
Then all components of the torsion vanish $\mathcal{T}_{\;\;\;ab}^{0}=\mathcal{T}_{\;\;\;ab}^{1}=\mathcal{T}_{\;\;\;ab}^{3}=0$
except for
\begin{eqnarray*}
\mathcal{T}_{\;\;\;ab}^{2} & = & \frac{\alpha}{2\kappa}\left(\begin{array}{cccc}
0 &  &  & D_{\bar{\mu}\mu}\\
 & 0\\
 &  & 0\\
-D_{\bar{\mu}\mu} &  &  & 0
\end{array}\right)
\end{eqnarray*}
Similarly, a spin up positron at rest with $\psi_{p0}=\left(0,0,0,\sigma\right)$
produces torsion given by $\mathcal{T}_{\;\;\;ab}^{0}=\mathcal{T}_{\;\;\;ab}^{1}=\mathcal{T}_{\;\;\;ab}^{3}=0$
and
\begin{eqnarray*}
\mathcal{T}_{\;\;\;ab}^{2} & = & \frac{\alpha}{2\kappa}\left(\begin{array}{cccc}
0 &  &  & -D_{\bar{\sigma}\sigma}\\
 & 0\\
 &  & 0\\
D_{\bar{\sigma}\sigma} &  &  & 0
\end{array}\right)
\end{eqnarray*}
These differ only by the replacement $\mu\rightarrow\sigma$ and an
overall sign. The simple relationship between particle and antiparticle
will be maintained by the $SO\left(3,1\right)$ subgroup since interchange
of past and future light cones is a Lorentz transformation.

Since nonmetricity breaks Lorentz invariance, the particle/antiparticle
cases may differ. However, the nonmetricity arising from a spin-up
electron is $\mathscr{Q}_{\quad ab}^{0}=\mathscr{Q}_{\quad ab}^{1}=\mathscr{Q}_{\quad ab}^{3}=0$
and
\begin{eqnarray*}
\mathscr{Q}_{\quad ab}^{2} & = & \frac{4\alpha}{\kappa}\left(\begin{array}{cccc}
0 &  &  & D_{\bar{\mu}\mu}\\
 & 0\\
 &  & 0\\
D_{\bar{\mu}\mu} &  &  & 0
\end{array}\right)
\end{eqnarray*}
while that of a spin-up positron at rest is similar, $\mathscr{Q}_{\quad ab}^{0}=\mathscr{Q}_{\quad ab}^{1}=\mathscr{Q}_{\quad ab}^{3}=0$,
and
\begin{eqnarray*}
\mathscr{Q}_{\quad ab}^{2} & = & \frac{4\alpha}{\kappa}\left(\begin{array}{cccc}
0 &  &  & D_{\bar{\sigma}\sigma}\\
 & 0\\
 &  & 0\\
D_{\bar{\sigma}\sigma} &  &  & 0
\end{array}\right)
\end{eqnarray*}

Other special cases follow easily.\medskip{}
\medskip{}

\begin{description}
\item [{{\large Acknowledgements}:}] The author thanks Matthew Pelligrini
and Joshua Knobloch for useful discussions.
\end{description}
\pagebreak{}

\pagebreak{}

\section*{Appendix: Spinor couplings \label{sec:Appendix-Spinor couplings}}

The interaction terms may be written as $S_{interaction}=-\alpha\int Re\left(i\psi^{\dagger}\left(h\gamma^{a}b_{Aa}\eta h\hat{\Gamma}^{A}\right)\psi\right)$.
When we expand the connection this gives 64 scalars, divided by symmetry,
24 for torsion sources and 40 for nonmetric sources. Here we list
the resulting scalars.

\subsection*{Spinor scalars for torsion}

The spinor scalars required for torsion sources are
\begin{eqnarray*}
\begin{array}{ccccccccc}
\frac{\alpha}{\kappa}Re\left(\psi^{\dagger}ih\gamma^{0}i\eta h\gamma^{2}\psi\right) & = & -\frac{i\alpha}{\kappa}\left(\bar{\nu}\rho-\bar{\rho}\nu\right) &  &  &  & \frac{\alpha}{\kappa}Re\left(\psi^{\dagger}ih\gamma^{2}i\eta h\gamma^{2}\psi\right) & = & \frac{\alpha}{\kappa}\left(-\bar{\mu}\mu-\bar{\nu}\nu-\bar{\rho}\rho+\bar{\sigma}\sigma\right)\\
\frac{\alpha}{\kappa}Re\left(\psi^{\dagger}ih\gamma^{0}\eta h\sigma^{01}\psi\right) & = & \frac{i\alpha}{\kappa}\left(\bar{\nu}\rho-\bar{\rho}\nu\right) &  &  &  & \frac{\alpha}{\kappa}Re\left(\psi^{\dagger}ih\gamma^{2}\eta h\sigma^{01}\psi\right) & = & \frac{\alpha}{\kappa}\left(-\bar{\mu}\mu+\bar{\nu}\nu+\bar{\rho}\rho+\bar{\sigma}\sigma\right)\\
\frac{\alpha}{\kappa}Re\left(\psi^{\dagger}ih\gamma^{0}i\eta h\gamma_{5}\gamma^{2}\psi\right) & = & -\frac{i\alpha}{\kappa}\left(\bar{\rho}\sigma-\bar{\sigma}\rho\right) &  &  &  & \frac{\alpha}{\kappa}Re\left(\psi^{\dagger}ih\gamma^{2}i\eta h\gamma_{5}\gamma^{2}\psi\right) & = & \frac{\alpha}{\kappa}\left(\bar{\mu}\rho+\bar{\rho}\mu\right)\\
\frac{\alpha}{\kappa}Re\left(\psi^{\dagger}ih\gamma^{0}\eta h\sigma^{31}\psi\right) & = & \frac{i\alpha}{\kappa}\left(\bar{\rho}\sigma-\bar{\sigma}\rho\right) &  &  &  & \frac{\alpha}{\kappa}Re\left(\psi^{\dagger}ih\gamma^{2}\eta h\sigma^{31}\psi\right) & = & -\frac{\alpha}{\kappa}\left(\bar{\nu}\sigma+\bar{\sigma}\nu\right)\\
\frac{\alpha}{\kappa}Re\left(\psi^{\dagger}ih\gamma^{0}\eta h\gamma_{5}\psi\right) & = & \frac{i\alpha}{\kappa}\left(\bar{\nu}\sigma-\bar{\sigma}\nu\right) &  &  &  & \frac{\alpha}{\kappa}Re\left(\psi^{\dagger}ih\gamma^{2}\eta h\gamma_{5}\psi\right) & = & \frac{\alpha}{\kappa}\left(\bar{\rho}\sigma+\bar{\sigma}\rho\right)\\
\frac{\alpha}{\kappa}Re\left(\psi^{\dagger}ih\gamma^{0}\eta h\sigma^{03}\psi\right) & = & -\frac{i\alpha}{\kappa}\left(\bar{\nu}\sigma-\bar{\sigma}\nu\right) &  &  &  & \frac{\alpha}{\kappa}Re\left(\psi^{\dagger}ih\gamma^{2}\eta h\sigma^{03}\psi\right) & = & \frac{\alpha}{\kappa}\left(\bar{\mu}\nu+\bar{\nu}\mu\right)\\
\\\frac{\alpha}{\kappa}Re\left(\psi^{\dagger}ih\gamma^{1}i\eta h\gamma^{2}\psi\right) & = & 0 &  &  &  & \frac{\alpha}{\kappa}Re\left(\psi^{\dagger}ih\gamma^{3}i\eta h\gamma^{2}\psi\right) & = & -\frac{i\alpha}{\kappa}\left(\bar{\rho}\sigma-\bar{\sigma}\rho\right)\\
\frac{\alpha}{\kappa}Re\left(\psi^{\dagger}ih\gamma^{1}\eta h\sigma^{01}\psi\right) & = & 0 &  &  &  & \frac{\alpha}{\kappa}Re\left(\psi^{\dagger}ih\gamma^{3}\eta h\sigma^{01}\psi\right) & = & -\frac{i\alpha}{\kappa}\left(\bar{\mu}\nu-\bar{\nu}\mu\right)\\
\frac{\alpha}{\kappa}Re\left(\psi^{\dagger}ih\gamma^{1}i\eta h\gamma_{5}\gamma^{2}\psi\right) & = & -\frac{i\alpha}{\kappa}\left(\bar{\nu}\sigma-\bar{\sigma}\nu\right) &  &  &  & \frac{\alpha}{\kappa}Re\left(\psi^{\dagger}ih\gamma^{3}i\eta h\gamma_{5}\gamma^{2}\psi\right) & = & -\frac{i\alpha}{\kappa}\left(\bar{\nu}\rho-\bar{\rho}\nu\right)\\
\frac{\alpha}{\kappa}Re\left(\psi^{\dagger}ih\gamma^{1}\eta h\sigma^{31}\psi\right) & = & -\frac{i\alpha}{\kappa}\left(\bar{\mu}\rho-\bar{\rho}\mu\right) &  &  &  & \frac{\alpha}{\kappa}Re\left(\psi^{\dagger}ih\gamma^{3}\eta h\sigma^{31}\psi\right) & = & \frac{i\alpha}{\kappa}\left(\bar{\mu}\sigma-\bar{\sigma}\mu\right)\\
\frac{\alpha}{\kappa}Re\left(\psi^{\dagger}ih\gamma^{1}\eta h\gamma_{5}\psi\right) & = & \frac{i\alpha}{\kappa}\left(\bar{\rho}\sigma-\bar{\sigma}\rho\right) &  &  &  & \frac{\alpha}{\kappa}Re\left(\psi^{\dagger}ih\gamma^{3}\eta h\gamma_{5}\psi\right) & = & 0\\
\frac{\alpha}{\kappa}Re\left(\psi^{\dagger}ih\gamma^{1}\eta h\sigma^{03}\psi\right) & = & \frac{i\alpha}{\kappa}\left(\bar{\mu}\nu-\bar{\nu}\mu\right) &  &  &  & \frac{\alpha}{\kappa}Re\left(\psi^{\dagger}ih\gamma^{3}\eta h\sigma^{03}\psi\right) & = & 0
\end{array}\\
\end{eqnarray*}

\subsection*{Spinor scalars for nonmetricity}

The spinor scalars required for nonmetricity sources are
\[
\begin{array}{ccccccc}
\alpha\delta b_{0}Re\left(\psi^{\dagger}ih\gamma^{0}\eta h\psi\right) & = & 0 &  & \alpha\delta b_{2}Re\left(\psi^{\dagger}ih\gamma^{2}\eta h\psi\right) & = & \alpha\delta b_{2}\left(\bar{\nu}\rho+\bar{\rho}\nu\right)\\
\alpha\delta b_{00}Re\left(\psi^{\dagger}ih\gamma^{0}\eta h\gamma^{0}\psi\right) & = & 0 &  & \alpha\delta b_{20}Re\left(\psi^{\dagger}ih\gamma^{2}\eta h\gamma^{0}\psi\right) & = & \alpha\delta b_{20}\left(\bar{\mu}\sigma+\bar{\sigma}\mu\right)\\
\alpha\delta b_{01}Re\left(\psi^{\dagger}ih\gamma^{0}\eta h\gamma^{1}\psi\right) & = & i\alpha\delta b_{01}\left(\bar{\sigma}\mu-\bar{\mu}\sigma\right) &  & \alpha\delta b_{21}Re\left(\psi^{\dagger}ih\gamma^{2}\eta h\gamma^{1}\psi\right) & = & \alpha\delta b_{21}\left(\bar{\mu}\mu-\bar{\nu}\nu+\bar{\rho}\rho+\bar{\sigma}\sigma\right)\\
\alpha\delta b_{03}Re\left(\psi^{\dagger}ih\gamma^{0}\eta h\gamma^{3}\psi\right) & = & -i\alpha\delta b_{03}\left(\bar{\mu}\rho-\bar{\rho}\mu\right) &  & \alpha\delta b_{23}Re\left(\psi^{\dagger}ih\gamma^{2}\eta h\gamma^{3}\psi\right) & = & -\alpha\delta b_{23}Re\left(\bar{\mu}\nu+\bar{\nu}\mu\right)\\
\alpha\delta b_{050}Re\left(\psi^{\dagger}ih\gamma^{0}\eta h\gamma_{5}\gamma^{0}\psi\right) & = & i\alpha\delta b_{050}\left(\bar{\mu}\rho-\bar{\rho}\mu\right) &  & \alpha\delta b_{250}Re\left(\psi^{\dagger}ih\gamma^{2}\eta h\gamma_{5}\gamma^{0}\psi\right) & = & -\alpha\delta b_{250}\left(\bar{\rho}\sigma+\bar{\sigma}\rho\right)\\
\alpha\delta b_{051}Re\left(\psi^{\dagger}ih\gamma^{0}\eta h\gamma_{5}\gamma^{1}\psi\right) & = & i\alpha\delta b_{051}\left(\bar{\mu}\nu-\bar{\nu}\mu\right) &  & \alpha\delta b_{251}Re\left(\psi^{\dagger}ih\gamma^{2}\eta h\gamma_{5}\gamma^{1}\psi\right) & = & -\alpha\delta b_{251}\left(\bar{\mu}\rho+\bar{\rho}\mu\right)\\
\alpha\delta b_{053}Re\left(\psi^{\dagger}ih\gamma^{0}\eta h\gamma_{5}\gamma^{3}\psi\right) & = & 0 &  & \alpha\delta b_{253}Re\left(\psi^{\dagger}ih\gamma^{2}\eta h\gamma_{5}\gamma^{3}\psi\right) & = & \alpha\delta b_{253}\left(\bar{\nu}\rho+\bar{\rho}\nu\right)\\
\alpha\delta b_{002}Re\left(\psi^{\dagger}ih\gamma^{0}i\eta h\sigma^{02}\psi\right) & = & -i\alpha\delta b_{002}\left(\bar{\mu}\sigma-\bar{\sigma}\mu\right) &  & \alpha\delta b_{202}Re\left(\psi^{\dagger}ih\gamma^{2}i\eta h\sigma^{02}\psi\right) & = & \alpha\delta b_{202}\left(\bar{\mu}\mu+\bar{\nu}\nu-\bar{\rho}\rho+\bar{\sigma}\sigma\right)\\
\alpha\delta b_{012}Re\left(\psi^{\dagger}ih\gamma^{0}i\eta h\sigma^{12}\psi\right) & = & 0 &  & \alpha\delta b_{212}Re\left(\psi^{\dagger}ih\gamma^{2}i\eta h\sigma^{12}\psi\right) & = & \alpha\delta b_{212}\left(\bar{\mu}\sigma+\bar{\sigma}\mu\right)\\
\alpha\delta b_{023}Re\left(\psi^{\dagger}ih\gamma^{0}i\eta h\sigma^{23}\psi\right) & = & -i\alpha\delta b_{023}\left(\bar{\mu}\nu-\bar{\nu}\mu\right) &  & \alpha\delta b_{223}Re\left(\psi^{\dagger}ih\gamma^{2}i\eta h\sigma^{23}\psi\right) & = & \alpha\delta b_{223}\left(\bar{\nu}\sigma+\bar{\sigma}\nu\right)\\
\\\alpha\delta b_{1}Re\left(\psi^{\dagger}ih\gamma^{1}\eta h\psi\right) & = & -i\alpha\delta b_{1}\left(\bar{\nu}\rho-\bar{\rho}\nu\right) &  & \alpha\delta b_{3}Re\left(\psi^{\dagger}ih\gamma^{3}\eta h\psi\right) & = & i\alpha\delta b_{3}\left(\bar{\nu}\sigma-\bar{\sigma}\nu\right)\\
\alpha\delta b_{10}Re\left(\psi^{\dagger}ih\gamma^{1}\eta h\gamma^{0}\psi\right) & = & i\alpha\delta b_{10}\left(\bar{\mu}\sigma-\bar{\sigma}\mu\right) &  & \alpha\delta b_{30}Re\left(\psi^{\dagger}ih\gamma^{3}\eta h\gamma^{0}\psi\right) & = & i\alpha\delta b_{30}\left(\bar{\mu}\rho-\bar{\rho}\mu\right)\\
\alpha\delta b_{11}Re\left(\psi^{\dagger}ih\gamma^{1}\eta h\gamma^{1}\psi\right) & = & 0 &  & \alpha\delta b_{31}Re\left(\psi^{\dagger}ih\gamma^{3}\eta h\gamma^{1}\psi\right) & = & i\alpha\delta b_{31}\left(\bar{\mu}\nu-\bar{\nu}\mu\right)\\
\alpha\delta b_{13}Re\left(\psi^{\dagger}ih\gamma^{1}\eta h\gamma^{3}\psi\right) & = & -i\alpha\delta b_{13}\left(\bar{\mu}\nu-\bar{\nu}\mu\right) &  & \alpha\delta b_{33}Re\left(\psi^{\dagger}ih\gamma^{3}\eta h\gamma^{3}\psi\right) & = & 0\\
\alpha\delta b_{150}Re\left(\psi^{\dagger}ih\gamma^{1}\eta h\gamma_{5}\gamma^{0}\psi\right) & = & -i\alpha\delta b_{150}\left(\bar{\rho}\sigma-\bar{\sigma}\rho\right) &  & \alpha\delta b_{350}Re\left(\psi^{\dagger}ih\gamma^{3}\eta h\gamma_{5}\gamma^{0}\psi\right) & = & 0\\
\alpha\delta b_{151}Re\left(\psi^{\dagger}ih\gamma^{1}\eta h\gamma_{5}\gamma^{1}\psi\right) & = & -i\alpha\delta b_{151}\left(\bar{\nu}\sigma-\bar{\sigma}\nu\right) &  & \alpha\delta b_{351}Re\left(\psi^{\dagger}ih\gamma^{3}\eta h\gamma_{5}\gamma^{1}\psi\right) & = & -i\alpha\delta b_{351}\left(\bar{\mu}\sigma-\bar{\sigma}\mu\right)\\
\alpha\delta b_{153}Re\left(\psi^{\dagger}ih\gamma^{1}\eta h\gamma_{5}\gamma^{3}\psi\right) & = & -i\alpha\delta b_{153}\left(\bar{\nu}\rho-\bar{\rho}\nu\right) &  & \alpha\delta b_{353}Re\left(\psi^{\dagger}ih\gamma^{3}\eta h\gamma_{5}\gamma^{3}\psi\right) & = & -i\alpha\delta b_{353}\left(\bar{\mu}\rho-\bar{\rho}\mu\right)\\
\alpha\delta b_{102}Re\left(\psi^{\dagger}ih\gamma^{1}i\eta h\sigma^{02}\psi\right) & = & 0 &  & \alpha\delta b_{302}Re\left(\psi^{\dagger}ih\gamma^{3}i\eta h\sigma^{02}\psi\right) & = & i\alpha\delta b_{302}\left(\bar{\rho}\sigma-\bar{\sigma}\rho\right)\\
\alpha\delta b_{112}Re\left(\psi^{\dagger}ih\gamma^{1}i\eta h\sigma^{12}\psi\right) & = & i\alpha\delta b_{112}\left(\bar{\mu}\sigma-\bar{\sigma}\mu\right) &  & \alpha\delta b_{312}Re\left(\psi^{\dagger}ih\gamma^{3}i\eta h\sigma^{12}\psi\right) & = & -i\alpha\delta b_{312}\left(\bar{\nu}\sigma-\bar{\sigma}\nu\right)\\
\alpha\delta b_{123}Re\left(\psi^{\dagger}ih\gamma^{1}i\eta h\sigma^{23}\psi\right) & = & -i\alpha\delta b_{123}\left(\bar{\mu}\rho-\bar{\rho}\mu\right) &  & \alpha\delta b_{323}Re\left(\psi^{\dagger}ih\gamma^{3}i\eta h\sigma^{23}\psi\right) & = & i\alpha\delta b_{323}\left(\bar{\nu}\rho-\bar{\rho}\nu\right)
\end{array}
\]

\end{document}